\documentclass[amssymb,preprint,showpacs,showkeys,nofootinbib,aps,floatfix]{revtex4}
\usepackage{psfig}
\usepackage{epsfig}
\usepackage{colordvi}
\usepackage{graphicx}
\def\slashed{{/}\mskip-10.0mu}

\def\pcirc{{p\mskip -5mu ^{^\circ}}}
\def\kcirc{{k\mskip -5mu ^{^\circ}}}
\def\qcirc{{q\mskip -5mu ^{^\circ}}}
\def\minuscirc{{-\mskip -10mu ^{^{^\circ}}}}
\newcommand{\real}{{\sf I}\kern-.12em{\sf R}}
\newcommand{\fillrhd}{{\rhd\kern-.81em\bullet}\kern-.47em\triangleright}
\begin{document}
\vskip 2cm

\title{Two-loop renormalization of vector, axial-vector and tensor
  fermion bilinears on the lattice} 

\author{A. Skouroupathis and H. Panagopoulos}
\affiliation{Department of Physics, University of Cyprus, P.O. Box 20537,
Nicosia CY-1678, Cyprus \\
{\it email: }{\tt php4as01@ucy.ac.cy, haris@ucy.ac.cy}}
\vskip 3mm

\date{\today}

\begin{abstract}

We compute the two-loop renormalization functions, in the $RI^\prime$
scheme, of local bilinear quark operators $\bar{\psi}\Gamma\psi$,
where $\Gamma$ corresponds to the Vector, Axial-Vector and Tensor
Dirac operators, in the lattice formulation of QCD. We consider both
the flavor nonsinglet and singlet operators.

We use the clover action for fermions and the Wilson action for
gluons. Our results are given as a polynomial in $c_{SW}$, in terms of
both the renormalized and bare coupling constant, in the renormalized
Feynman gauge.  

Finally, we present our results in the $\overline{MS}$ scheme, for
easier comparison with calculations in the continuum. The
corresponding results, for fermions in an {\it arbitrary}
representation, together with some special features of superficially 
divergent integrals, are included in the Appendices. 

\end{abstract}

\medskip
\medskip

\maketitle

\newpage


\section{Introduction}
\label{introduction}

Numerical simulations of QCD, formulated on the lattice, make use of a
variety of composite operators, made out of quark fields.
In particular, matrix elements and
correlation functions of such operators, which include local and
extended bilinears, as well as four-fermion operators, are computed in order to
study hadronic properties in this context. A proper renormalization of these
operators is essential for the extraction of physical results from the
dimensionless quantities measured in numerical simulations.

The present paper is the second in a series of papers regarding the
calculation of renormalization functions of fermion bilinear operators
to two loops in Lattice perturbation theory. The calculation of the scalar and
pseudoscalar cases was carried out in Ref. \cite{SP-08}. In this work
we study the renormalization function $Z_\Gamma$ of fermion bilinears
${\cal O}=\bar{\psi}\Gamma\psi$ on the lattice, where
$\Gamma=\gamma_{\mu},\,\gamma_5\,\gamma_{\mu},\,\gamma_5\,\sigma_{\mu\,\nu}$
($\sigma_{\mu\,\nu}=1/2\,[\gamma_\mu,\gamma_\nu]$). We consider both
flavor singlet and nonsinglet operators. We employ the standard Wilson
action for gluons and clover-improved Wilson fermions. The number of
quark flavors $N_f$, the number of colors $N_c$ and the clover
coefficient $c_{{\rm SW}}$ are kept as free parameters. One necessary
ingredient for the renormalization of fermion bilinears is the 2-loop
quark field renormalization, $Z_{\psi}$, calculated in
\cite{SP-08}. The one-loop expression for the renormalization function
$Z_g$ of the coupling constant is also necessary for expressing the
results in terms of both the bare and the renormalized coupling constant.   

Our two-loop calculations have been performed in the bare and in 
the renormalized Feynman gauge. For the latter, we need the 1-loop 
renormalization functions $Z_\alpha$ and $Z_A$ of the gauge
parameter and gluon field respectively, as well as the one-loop
expressions for $Z_\Gamma$ with an arbitrary value of the gauge
parameter. 
 
The main results presented in this work are the following 
2-loop bare Green's functions (amputated, one-particle irreducible (1PI)): 
\begin{itemize}
\item 2-pt function of the vector operator $\bar{\psi}\gamma_\mu\psi:
$ $\Sigma^L_V(q a_{_{\rm L}})$ 
\item 2-pt function of the axial-vector operator $\bar{\psi}\gamma_5\gamma_\mu\psi:
$ $\Sigma^L_{AV}(q a_{_{\rm L}})$  
\item 2-pt function of the tensor operator $\bar{\psi}\gamma_5\sigma_{\mu\,\nu}\psi:
$ $\Sigma^L_T(q a_{_{\rm L}})$  
\end{itemize}
($a_{_{\rm L}}\,:$ lattice spacing, $q:$ external momentum).

In general, one can use bare Green's functions to construct 
$Z_{{\cal O}}^{X,Y}$, the renormalization function for 
operator ${\cal O}$, computed within a regularization $X$ ($X=L$:
lattice regularization; $X=DR$: dimensional regularization) and
renormalized in a scheme $Y$. We employ two widely used schemes to
compute the various two-loop renormalization functions:  
\begin{itemize}
\item The $RI^{\prime}$ scheme: $Z_V^{L,RI^{\prime}}$, 
$Z_{AV}^{L,RI^{\prime}}$, $Z_T^{L,RI^{\prime}}$ 
\item The $\overline{MS}$ scheme: $Z_V^{L,\overline{MS}}$, 
$Z_{AV}^{L,\overline{MS}}$, $Z_T^{L,\overline{MS}}$
\end{itemize}

For convenience, the results for $Z_{{\cal O}}^{X,Y}$
are given in terms of both the bare coupling
constant $g_{\rm o}$ and the renormalized one: $g_{RI^{\prime}}$, 
$g_{\overline{MS}}$. Finally, as one of several checks on our results,
we construct the 2-loop renormalized Green's functions in  $RI^{\prime}$: 
$\Sigma_{{\cal O}}^{RI^{\prime}}(q,\bar{\mu})$ (${\cal O}\equiv V,AV,T$),
as well as their counterparts in $\overline{MS}$:
$\Sigma_{{\cal O}}^{\overline{MS}}(q,\bar{\mu})$.
The values of all these functions, computed on the lattice, 
coincide with values computed in dimensional regularization
(we derive the latter from the results of Ref. \cite{Gracey}).

The present work, along with \cite{SP-08}, is the first two-loop
computation of the renormalization of fermion bilinears on the
lattice. One-loop computations of the same quantities exist for quite
some time now (see, e.g., \cite{MZ}, \cite{Aoki98}, \cite{Capitani}
and references therein). There have been made several attempts to
estimate $Z_{{\cal O}}$ non-perturbatively; recent results can be
found in 
Refs. \cite{Zhang,Becirevic,Aoki,Galletly,Sommer,DellaMorte}. 
A series of results have also been obtained using stochastic perturbation
theory \cite{Miccio,DiRenzo,Scorzato}. A related computation, regarding the fermion
mass renormalization $Z_m$ with staggered fermions can be found in
\cite{Trottier}. 

The paper is organized as follows: Section \ref{Formulation} provides
a formulation of the problem, as well as all necessary definitions of 
renormalization schemes and of the quantities to compute. Section 
\ref{Results} describes our computational methods and
results. Finally, in Section \ref{Discussion} we discuss some salient 
features of our calculation, and comment on future extensions to the
present work.  

Recently, there has been some interest in gauge theories with
fermions in different representations \cite{Sannino1} of the gauge group. Such
theories are being studied in various contexts
\cite{KUY,GG,CDDLP,BPV,GK,EHS}, e.g., supersymmetry \cite{Sannino2}, phase
transitions \cite{Sannino3}, and the AdS/QCD correspondence. It is relatively
straightforward to generalize our results to an arbitrary
representation; this is presented in Appendix A. Some special features
of 2-, 3- and 4-index superficially divergent integrals are described
in Appendix B. Finally, a detailed presentation of our calculation
results on a per diagram basis, is provided in Appendix C. 

\newpage

\section{Formulation of the problem}
\label{Formulation}

\subsection{Lattice action} 

In the present work we employ the Wilson formulation of the QCD action 
on the lattice, with the addition of the clover (SW) \cite{SW} term 
for fermions. In standard notation, it reads:  

\begin{eqnarray}
S_L &=& S_G + \sum_{f}\sum_{x} (4r+m_{\rm o})\bar{\psi}_{f}(x)\psi_f(x)
\nonumber \\
&-& {1\over 2}\sum_{f}\sum_{x,\,\mu}
\bigg{[}\bar{\psi}_{f}(x) \left( r - \gamma_\mu\right)
U_{x,\,x+\mu}\,\psi_f(x+{\mu}) \nonumber \\
&~& \hspace{1.7cm}+\bar{\psi}_f(x+{\mu})\left( r + \gamma_\mu\right)
U_{x+\mu,\,x}\,\psi_{f}(x)\bigg{]}\nonumber \\
&+& {i\over 4}\,c_{\rm SW}\,\sum_{f}\sum_{x,\,\mu,\,\nu} \bar{\psi}_{f}(x)
\,\sigma_{\mu\nu} \,{\hat F}_{\mu\nu}(x) \,\psi_f(x),
\label{latact}
\end{eqnarray}
\begin{eqnarray}
{\rm where:}\qquad {\hat F}_{\mu\nu} &\equiv& {1\over{8a^2}}\,
(Q_{\mu\nu} - Q_{\nu\mu})\\
{\rm and:\qquad} Q_{\mu\nu} &=& U_{x,\, x+\mu}\,U_{x+\mu,\, x+\mu+\nu}\,U_{x+\mu+\nu,\, x+\nu}
\,U_{x+\nu,\, x}\nonumber \\
&+& U_{ x,\, x+ \nu}\,U_{ x+ \nu,\, x+ \nu- \mu}\,U_{ x+ \nu- \mu,\, x- \mu}\,U_{ x- \mu,\, x} \nonumber \\
&+& U_{ x,\, x- \mu}\,U_{ x- \mu,\, x- \mu- \nu}\,U_{ x- \mu- \nu,\, x- \nu}\,U_{ x- \nu,\, x}\nonumber \\
&+& U_{ x,\, x- \nu}\,U_{ x- \nu,\, x- \nu+ \mu}\,U_{ x- \nu+ \mu,\, x+ \mu}\,U_{ x+ \mu,\, x}
\label{latact2}
\end{eqnarray}

$S_G$ is the standard pure gluon action, made out of $1{\times}1$
plaquettes. The clover coefficient $c_{\rm SW}$ is treated here as a
free parameter; $r$ is the Wilson parameter (set to $r=1$ henceforth); 
$f$ is a flavor index; $\sigma_{\mu\nu} =(i/2) [\gamma_\mu,\,\gamma_\nu]$.
Powers of the lattice spacing $a_{_{\rm L}}$ have been omitted and may
be directly reinserted by dimensional counting.  

The ``Lagrangian mass'' $m_{\rm o}$ is a free parameter here. However,
since we will be using mass independent renormalization schemes, all
renormalization functions which we will be calculating, must be
evaluated at vanishing renormalized mass, that is, when $m_{\rm o}$ is
set equal to the critical value $m_{\rm cr}$: $m_{\rm o}\to m_{\rm
  cr}=m_1\,g_\circ^2+{\cal O}(g_\circ^4)$. 

\subsection{Definition of renormalized quantities}

As a prerequisite to our programme, we will use the renormalization
functions, $Z_A$, $Z_c$, $Z_\psi$, $Z_g$ and $Z_\alpha$, for the
gluon, ghost and fermion fields ($A_\mu^a,\ c^a,\ \psi$), and for the
coupling constant $g$ and gauge parameter $\alpha$, respectively (for
definitions of these quantities, see Ref. \cite{SP-08}); we will also
need the fermion mass counterterm $m_{\rm cr}$. These quantities are
all needed to one loop, except for $Z_\psi$ which is required to two
loops.  The value of each $Z_{{\cal O}}$ depends both on the
regularization $X$ and on the renormalization scheme $Y$ employed, and
thus should properly be denoted as $Z^{X,Y}_{{\cal O}}$. 

Our one-loop results for the Vector and Axial-Vector operators, even
though performed in a generic gauge, turn out to be independent of the
gauge parameter. These results along with the one-loop expression for
the Tensor operator, are in agreement with results found in the
literature (see, e.g., Ref. \cite{Capitani}). 

As mentioned before, we employ the $RI^\prime$ renormalization scheme
\cite{Martinelli,Franco,Chetyrkin2000}, which is more immediate for a
lattice regularized theory. It is defined by imposing a set of
normalization conditions on matrix elements at a scale $\bar{\mu}$,
where (just as in the $\overline{MS}$ scheme) \cite{Collins}: 

\begin{equation}
\bar{\mu}=\mu\,(4\pi/e^{\gamma_{\rm E}})^{1/2} 
\label{mubar}
\end{equation}
where $\gamma_{{\rm E}}$ is the Euler constant and $\mu$ is the scale
entering the bare coupling constant $g_\circ = \mu^\epsilon\,Z_g\,g$
when regularizing in $D=4-2\epsilon$ dimensions.

\subsection{Conversion to the $\mathbf{\overline{MS}}$ scheme}
 
For easier comparison with calculations coming from the continuum, we need
to express our results in the $\overline{MS}$ scheme. Each
renormalization function on the lattice, $Z^{L,RI^{\prime}}_{{\cal O}}$, 
may be expressed as a power series in the renormalized
coupling constant $g_{RI^{\prime}}$. 
For the purposes of our work the conversion of $g_{RI^{\prime}}$ to 
$\overline{MS}$ is trivial since:

\begin{equation}
g_{RI^{\prime}}=g_{\overline{MS}}+{\cal O}\left((g_{\overline{MS}})^9\right) 
\label{gConversion}
\end{equation}

The conversion of the gauge parameter $\alpha_{RI^\prime}$ to the
$\overline{MS}$ scheme is given by \cite{Retey}: 

\begin{equation}
\alpha_{RI^{\prime}}=\frac{Z_A^{L,\overline{MS}}}{Z_A^{L,RI^{\prime}}}\,\alpha_{\overline{MS}}
\equiv \alpha_{\overline{MS}}\, /\, C_A(g_{\overline{MS}}, \alpha_{\overline{MS}})
\label{alphaConversion}
\end{equation}
where the conversion factor $C_A$ may be calculated more easily in
dimensional regularization (DR) \cite{Gracey}, since the ratio of $Z$'s
appearing in Eq. (\ref{alphaConversion}) is necessarily {\em regularization
  independent}. To one loop, the conversion factor  $C_A$ equals: 
\begin{equation}
C_A(g,\alpha)=\frac{Z_A^{DR,RI^{\prime}}}{Z_A^{DR,\overline{MS}}}=1+\frac{g^2}{36(16\pi^2)}\,\left[\left(9\alpha^2+ 
18\alpha +97\right)\,N_c - 40N_f\right]
\label{CA}
\end{equation}
(Here, and throughout the rest of this work, both $g$ and $\alpha$ are 
in the $\overline{MS}$ scheme, unless specified otherwise.) 

Once we have computed the renormalized Green's functions in the
$RI^{\prime}$ scheme, we can construct their $\overline{MS}$
counterparts using the quark field conversion factor which, up to the
required perturbative order, is given by:

\begin{eqnarray}
C_{\psi}(g,\alpha)&\equiv&\frac{Z_{\psi}^{L,RI^{\prime}}}{Z_{\psi}^{L,\overline{MS}}}=\frac{Z_{\psi}^{DR,RI^{\prime}}}{Z_{\psi}^{DR,\overline{MS}}} \nonumber \\
&=& 1 - \frac{g^2}{16\pi^2}\,\,c_F\,\alpha 
+\frac{g^4}{8\,(16\pi^2)^2}\,c_F\bigg[\left(8\alpha^2 + 5\right)\,c_F +  14\,N_f \nonumber \\ 
&& \hskip 5.2cm -\left(9\alpha^2-24\zeta(3)\,\alpha+52\alpha-24\zeta(3)+82\right)\,N_c\bigg] 
\label{Cpsi}
\end{eqnarray}
where $c_F= (N_c^2-1)/(2\,N_c)$ is the quadratic Casimir operator in
the fundamental representation of the color group; $\zeta(x)$ is
Riemann's zeta function. 

\subsection{Renormalization of fermion bilinears} 

The lattice operators ${\cal O}_{\Gamma}=\bar{\psi}\,\Gamma\,\psi$
must, in general, be renormalized in order to have finite matrix
elements. We define renormalized operators by 
\begin{equation}
{\cal
  O}^{RI^{\prime}}_{\Gamma}=Z^{L,RI^{\prime}}_{\Gamma}(a_{_{\rm L}}\bar\mu)\,{\cal
  O}_{\Gamma\,{\rm o}}
\label{RenormOper}
\end{equation}

The flavor singlet axial-vector Green's function receives additional
contributions as compared to the nonsinglet case,
while for the rest of the operators under study, singlet and
nonsinglet Green's functions coincide. For the vector (V), axial-vector (AV)
and tensor (T) operators, the renormalization functions
$Z_{\Gamma}^{L,RI^{\prime}}$ can be extracted through the corresponding
bare 2-point functions $\Sigma^L_{\Gamma}(q a_{_{\rm L}})$ (amputated,
1PI) on the lattice. Let us first express these bare Green's functions 
in the following way:

\begin{eqnarray}
\Sigma^L_V(q a_{_{\rm L}})&=&\gamma_\mu\,\Sigma^{(1)}_V (q a_{_{\rm L}}) +
\frac{q^\mu\slashed q}{q^2}\Sigma^{(2)}_V (q a_{_{\rm L}}) \nonumber \\ 
\Sigma^L_{AV}(q a_{_{\rm L}})&=&\gamma_5\gamma_\mu\,\Sigma^{(1)}_{AV} (q a_{_{\rm L}}) + 
\gamma_5\frac{q^\mu\slashed q}{q^2}\Sigma^{(2)}_{AV} (q a_{_{\rm L}}) \label{2ptFunct}
\\
\Sigma^L_T(q a_{_{\rm L}})&=&\gamma_5\,\sigma_{\mu\,\nu}\Sigma^{(1)}_T(q a_{_{\rm L}}) + 
\gamma_5\frac{\slashed q\,(\gamma_\mu q_\nu - \gamma_\nu
  q_\mu)}{q^2}\Sigma^{(2)}_T(q a_{_{\rm L}})  \nonumber 
\label{brgrnfunctn}
\end{eqnarray}
It is worth noting here that terms which break Lorentz invariance (but
are compatible with hypercubic invariance), such as
$\gamma_\mu\,(q^\mu)^2/q^2$, turn out to be absent from all bare
Green's functions; thus, the latter have the same Lorentz structure as
in the continuum. Let us also point out that the presence of the
$\gamma_5$ matrix in the tensor operator definition does not affect
the bare Green's function on the lattice, in the $RI'$ scheme. We have
performed the calculation both with and without the inclusion of the
$\gamma_5$ matrix, and we ended up with identical 2-point
functions. Thus, for the purpose of converting our results to the
$\overline{MS}$ scheme, we employed the conversion factors given in
Ref. \cite{Gracey}, where the definition of the tensor operator does
not contain the $\gamma_5$ matrix. Furthermore, we expect that
$\Sigma^{(2)}_T(q a_{_{\rm L}})$ must vanish, since this is the case
for the corresponding quantity coming from the continuum. Indeed,
after performing the calculation on the lattice, it turns out that all
contributions of this type vanish.  

Once all necessary Feynman diagrams contributing to the
bare Green's functions presented above are evaluated, one can obtain
the renormalization functions for the three operators through the
following conditions: 

\begin{eqnarray}
\lim_{a_{_{\rm L}}\rightarrow 0}\left[Z_{\psi}^{L,RI^{\prime}}\,Z_V^{L,RI^{\prime}}\,\Sigma^{(1),\,L}_V (q a_{_{\rm L}})\right]_{q^2=\bar{\mu}^2} &=& \gamma_\mu \label{ZXrulesa} \\
\lim_{a_{_{\rm L}}\rightarrow 0}\left[Z_{\psi}^{L,RI^{\prime}}\,Z_{AV}^{L,RI^{\prime}}\,\Sigma^{(1),\,L}_{AV} (q a_{_{\rm L}})\right]_{q^2=\bar{\mu}^2} &=& \gamma_5\,\gamma_\mu
\label{ZXrulesb} \\
\lim_{a_{_{\rm L}}\rightarrow 0}\left[Z_{\psi}^{L,RI^{\prime}}\,Z_T^{L,RI^{\prime}}\,\Sigma^{(1),\,L}_T (q a_{_{\rm L}})\right]_{q^2=\bar{\mu}^2} &=& \gamma_5\,\sigma_{\mu\,\nu}
\label{ZXrulesc}
\end{eqnarray}
where:
\begin{eqnarray}
\Sigma^{(1)}_V (q a_{_{\rm L}}) &=& 1 + {\cal O}(g_\circ^2)\quad,\quad
\Sigma^{(2)}_V (q a_{_{\rm L}}) = {\cal O}(g_\circ^2) \nonumber \\
\Sigma^{(1)}_{AV} (q a_{_{\rm L}}) &=& 1 + {\cal O}(g_\circ^2)\quad,\quad 
\Sigma^{(2)}_{AV} (q a_{_{\rm L}}) = {\cal O}(g_\circ^2)  \\
\Sigma^{(1)}_T (q a_{_{\rm L}}) &=& 1 + {\cal O}(g_\circ^2)\quad,\quad
\Sigma^{(2)}_T (q a_{_{\rm L}}) = 0 \nonumber 
\label{2ptFunctbreakdown}
\end{eqnarray}

The conversion of the quantities $Z_\Gamma^{L,RI^{\prime}}$ to the
$\overline{MS}$ scheme is a straightforward procedure. In  
the case of the vector and tensor operators, the renormalization
functions, $Z_V^{L,\overline{MS}}$ and $Z_T^{L,\overline{MS}}$, can be
obtained by:  
\begin{equation}
Z_\Gamma^{L,\overline{MS}}=Z_\Gamma^{L,RI^{\prime}} / C_\Gamma(g,\alpha)
\label{VTConversion} 
\end{equation}
where $C_\Gamma(g,\alpha)$ are {\em regularization independent}
conversion factors ($\Gamma = V,T$). These conversion factors have
been calculated in dimensional regularization \cite{Gracey}: 
\begin{eqnarray}
C_V(g,\alpha)&\equiv&\frac{Z_V^{L,RI^{\prime}}}{Z_V^{L,\overline{MS}}}=
\frac{Z_V^{DR,RI^{\prime}}}{Z_V^{DR,\overline{MS}}} = 1+{\cal
  O}(g^8) 
\label{CV}\\
C_T (g,\alpha)&\equiv&\frac{Z_T^{L,RI^{\prime}}}{Z_T^{L,\overline{MS}}}=
\frac{Z_T^{DR,RI^{\prime}}}{Z_T^{DR,\overline{MS}}} \nonumber \\
&=& 1 + \frac{g^2}{16\pi^2}\,c_F\,\alpha + \frac{g^4}{216\,(16\pi^2)^2}\,c_F\,\bigg[\left(216\alpha^2+4320\zeta(3)-4815\right)\,c_F - 626\,N_f \nonumber \\
&& \hskip 5.4cm + \left(162\alpha^2+756\alpha -3024\zeta(3)+5987\right)\,N_c\bigg] 
\label{CT}
\end{eqnarray}

Unlike the tensor operator, where the presence of the $\gamma_5$ matrix
is irrelevant, the axial-vector bilinear (${\cal
  O}_{AV}^\circ=\bar{\psi}_\circ\gamma_5\gamma_\mu\psi_\circ$) requires
special attention also in the $\overline{MS}$ scheme, due to the
non-unique generalization of $\gamma_5$ to D dimensions. A practical
definition of $\gamma_5$ for multiloop calculations, which is most
commonly employed in dimensional regularization and does not suffer
from inconsistencies is \cite{Veltman}:  

\begin{equation}
\gamma_5=i\,\frac{1}{4!}\,\epsilon^{\nu_1\,\nu_2\,\nu_3\,\nu_4}\,\gamma_{\nu_1}\,\gamma_{\nu_2}\,\gamma_{\nu_3}\,\gamma_{\nu_4}
\quad,\quad \nu_i=0,\,1,\,2,\,3
\label{gamma5}
\end{equation}

Of course, $\gamma_5$ as defined in Eq. (\ref{gamma5}) does not
anticommute (in $D$ dimensions) with $\gamma_{\mu}$, for $\mu \geq 4$;
an ultimate consequence of this fact is that Ward identities involving
the axial-vector and pseudoscalar operators, renormalized in this way,
are violated.   

To obtain a correctly normalized axial-vector operator \cite{Larin},
${\cal O}_{AV}^{\overline{MS}^{\,\prime}}$, one must introduce an extra {\em 
  finite} factor, $Z_5$, in addition to the usual renormalization
function $Z^{DR,\overline{MS}}_{AV}$ (the latter only contains poles in
$\epsilon$). We set: 

\begin{equation}
{\cal O}_{AV}^{\overline{MS}^{\,\prime}}= Z_5(g)\,{\cal O}_{AV}^{\overline{MS}}=Z_5(g)\,Z^{DR,\overline{MS}}_{AV}\,{\cal O}_{AV}^\circ
\label{axialCurrent}
\end{equation}

For the definition of $Z_5$ we must express the $\overline{MS}$
renormalized Green's functions $G_V^{\overline{MS}}$,
$G_{AV}^{\overline{MS}}$ as well as the renormalized Green's function
$G_{AV}^{\overline{MS}^{\,\prime}}$ (corresponding to ${\cal
  O}_{AV}^{\overline{MS}^{\,\prime}}$), in a form similar to
Eq. (\ref{brgrnfunctn}): 
\begin{eqnarray}
G^{\overline{MS}}_V(q a_{_{\rm L}})&=&\gamma_\mu\,G^{(1)\,\overline{MS}}_V (q a_{_{\rm L}}) +
\frac{q^\mu\slashed q}{q^2}G^{(2)\,\overline{MS}}_V (q a_{_{\rm L}}) \nonumber \\ 
G^{\overline{MS}}_{AV}(q a_{_{\rm L}})&=&\gamma_5\gamma_\mu\,G^{(1)\,\overline{MS}}_{AV} (q a_{_{\rm L}}) + 
\gamma_5\frac{q^\mu\slashed q}{q^2}G^{(2)\,\overline{MS}}_{AV} (q a_{_{\rm L}}) \label{rngrnfunctn}
\\
G^{\overline{MS}^{\,\prime}}_{AV}(q a_{_{\rm L}})&=&\gamma_5\,\sigma_{\mu\,\nu}G^{(1)\,\overline{MS}^{\,\prime}}_{AV}(q a_{_{\rm L}}) + 
\gamma_5\frac{\slashed q\,(\gamma_\mu q_\nu - \gamma_\nu
  q_\mu)}{q^2}G^{(2)\,\overline{MS}^{\,\prime}}_{AV}(q a_{_{\rm L}})  \nonumber 
\end{eqnarray}
$Z_5$ is then defined by the requirement that the renormalized Green's
functions $G_V^{(1)\,\overline{MS}}(q a_{_{\rm L}})$ and
  $G_{AV}^{(1)\,\overline{MS}^{\,\prime}}(q a_{_{\rm L}})$ coincide: 
\begin{equation}
Z_5\equiv\frac{G_V^{(1)\,\overline{MS}}}{G_{AV}^{(1)\,\overline{MS}}}
\label{Z5def}
\end{equation}
Eq.(\ref{axialCurrent}) is valid for both the singlet and nonsinglet
currents, provided of course, the appropriate choice for $Z_5$ is
used. Thus, we have two different expressions, $Z_5^s$ and $Z_5^{ns}$
corresponding to the singlet and nonsinglet axial-vector operator,
respectively. They are gauge independent and differ only in the
$c_F\,N_f$ term; this is expected considering the fact that the
additional Feynman diagrams contributing to the singlet axial operator
have the insertion within the closed fermion loop. Both $Z_5^s$ and
$Z_5^{ns}$ were evaluated in DR \cite{Larin} and up to two loops they read:
\begin{eqnarray}
Z_5^s(g)&=&1-\frac{g^2}{16\pi^2}\,(4\,c_F)+\frac{g^4}{(16\pi^2)^2}\,
\left(22\,c_F^2 - \frac{107}{9}\,c_F\,N_c+\frac{31}{18}\,c_F\,N_f\right)
\label{Z5s} \\
Z_5^{ns}(g)&=&1-\frac{g^2}{16\pi^2}\,(4\,c_F)+\frac{g^4}{(16\pi^2)^2}\,
\left(22\,c_F^2 - \frac{107}{9}\,c_F\,N_c+\frac{2}{9}\,c_F\,N_f\right)
\label{Z5ns} 
\end{eqnarray}
$Z_{AV}^{L,\overline{MS}}$ can now be obtained by:
\begin{equation}
Z_{AV}^{L,\overline{MS}} = Z_{AV}^{L,RI^{\prime}} / \left( C_V\,Z_5 \right)
\label{AxialConversion}
\end{equation}
where $Z_5$ stands for $Z^s_5$ or $Z^{ns}_5$
(Eqs. (\ref{Z5s}-\ref{Z5ns})), for the singlet or nonsinglet cases,
respectively. 

Similarly, one can convert the $RI^{\prime}$ renormalized Green's functions,
$G_\Gamma^{RI^{\prime}}$, to their $\overline{MS}$ counterparts, through:

\begin{equation}
\frac{G^{RI^{\prime}}_V}{G^{\overline{MS}}_V} = C_{\psi}\,C_V \quad ,\quad
\frac{G^{RI^{\prime}}_{AV}}{G^{\overline{MS}}_{AV}} = C_{\psi}\,C_V\,Z_5
\quad ,\quad
\frac{G^{RI^{\prime}}_T}{G^{\overline{MS}}_T} = C_{\psi}\,C_T 
\label{RenormGreen}
\end{equation}
(In Eqs.(\ref{AxialConversion}-\ref{RenormGreen}) it is understood
that powers of $g_{RI^{\prime}},\ \alpha_{RI^{\prime}}$, implicit in
$RI^{\prime}$ quantities, must also be converted to 
$g_{\overline{MS}},\ \alpha_{\overline{MS}}$, respectively, using
Eqs.(\ref{gConversion}-\ref{alphaConversion})). Note that the
combination $C_V\,Z_5$ appearing above yields the value of
$C_{AV}\equiv
Z_{AV}^{DR,RI^\prime}/Z_{AV}^{DR,\overline{MS}}=C_V\,Z_5$.  

\newpage

\section{Computation and Results}
\label{Results}

\begin{figure}[b]
\centerline{\includegraphics[scale=0.40]{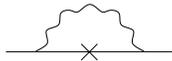}}
\caption{One-loop diagram contributing to $Z_V$, $Z_{AV}$ and $Z_T$. A
  wavy (solid) line represents gluons (fermions). A cross denotes the
  Dirac matrices $\gamma_\mu$ (vector), $\gamma_5\gamma_\mu$
  (axial vector) and $\gamma_{5}\sigma_{\mu\nu}$
  (tensor). \label{ZVAT1loopDiagrams}}  
\end{figure}

\begin{figure}
\centerline{\psfig{figure=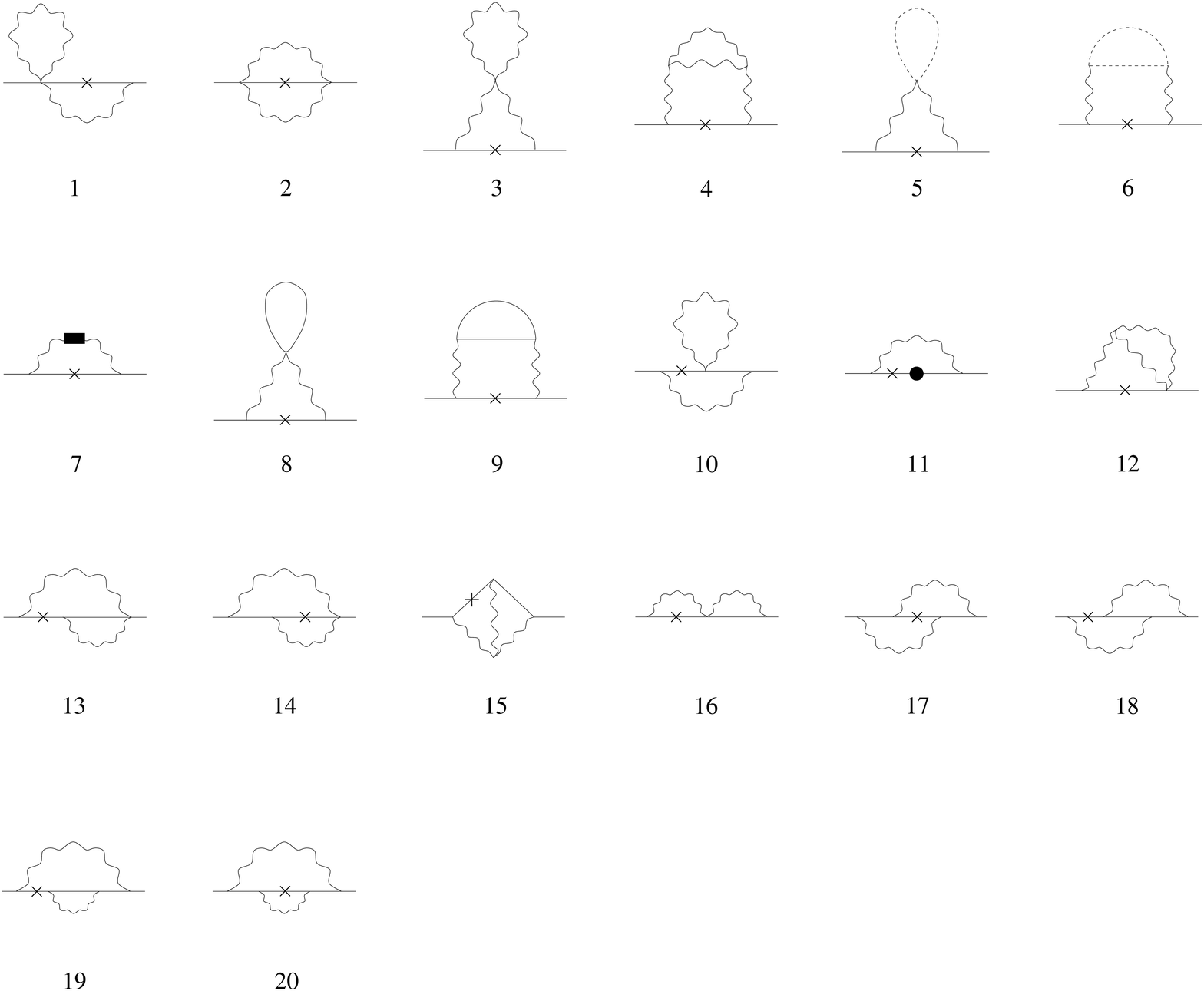,scale=0.35}}
\caption{Two-loop diagrams contributing to $Z_V$, $Z_{AV}$ and $Z_T$. 
  Wavy (solid, dotted) lines represent gluons (fermions, ghosts). A
  solid box denotes a vertex from the measure part of the action; a
  solid circle is a mass counterterm; crosses denote the matrices
  $\gamma_\mu$ (vector), $\gamma_5\gamma_\mu$ (axial-vector) and
  $\gamma_{5}\sigma_{\mu\nu}$ (tensor). \label{ZVAT2loopDiagrams}}  
\end{figure}

\begin{figure}
\centerline{\psfig{figure=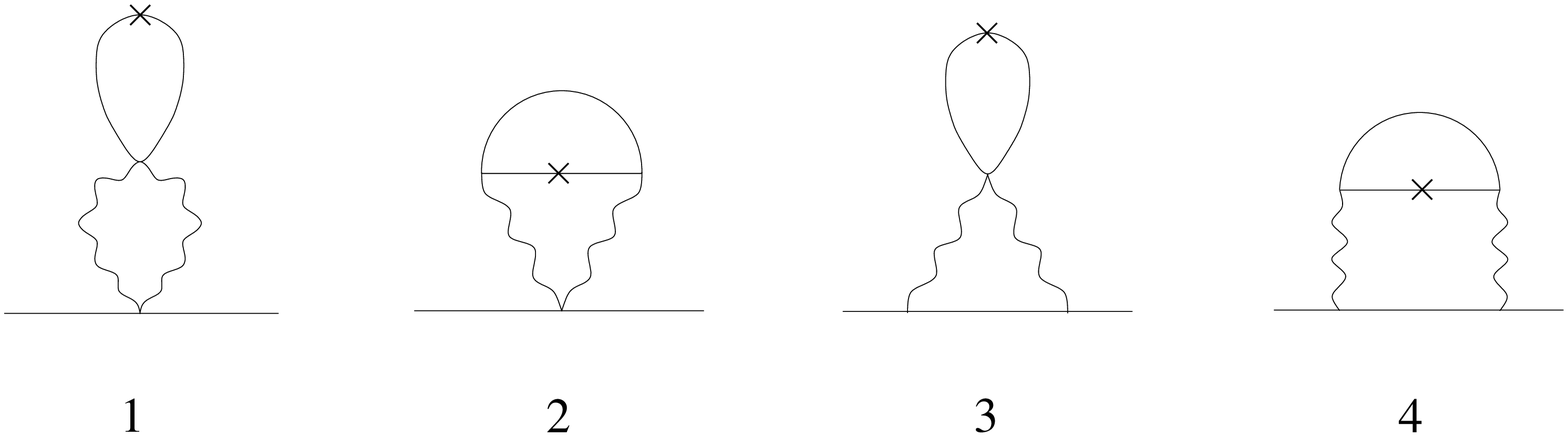,height=3.0truecm}}
\caption{Extra two-loop diagrams contributing to $Z_{AV,\,singlet}$. A
  cross denotes an insertion of a flavor singlet operator. Wavy
  (solid) lines represent gluons (fermions). \label{singletVAT}}  
\end{figure}

The Feynman diagrams contributing to the bare Green's functions for
the vector, axial-vector and tensor operators,
$\Sigma^L_{V,AV,T}(q,a_{_{\rm L}})$, at 1- and 2-loop level, are shown
in Figs. \ref{ZVAT1loopDiagrams} and \ref{ZVAT2loopDiagrams},
respectively. For flavor singlet bilinears, there are 4 extra
diagrams, shown in Fig. \ref{singletVAT}, which contain the operator
insertion inside a closed fermion loop. These diagrams give a nonzero
contribution only in the axial-vector case. 

The evaluation and algebraic manipulation of Feynman diagrams, leading
to a code for numerical loop integration, is performed automatically
using our software for Lattice Perturbation Theory, written in
Mathematica. 

The most laborious aspect of the procedure is the extraction of the
dependence on the external momentum $q$. This is a delicate task at
two loops; for this purpose, we cast algebraic expressions (typically
involving thousands of summands) into terms which can be naively
Taylor expanded in $q$ to the required order, plus a smaller set of
terms containing superficial divergences and/or subdivergences. The
latter can be evaluated by an extension of the method of
Ref. \cite{KNS} to 2 loops; this entails analytical continuation to
$D>4$ dimensions, and splitting each expression into a UV-finite part
(which can thus be calculated in the continuum, using the methods of
Ref. \cite{Chetyrkin}), and a part which is polynomial in $q$. A
primitive set of divergent lattice integrals involving gluon
propagators, which can be obtained in this manner, can be found in
Ref. \cite{Luscher}. Due to the presence of at least one free Lorentz
index in the definition of the operators (for the case of the tensor
bilinear there are two such indices), it is possible to end up dealing
with superficially divergent integrals with two, three or even four
free Lorentz indices. In Appendix B, we provide a brief description of
the manipulations performed to resolve such terms, based on the method
described above. 

Some of the diagrams contributing to $\Sigma^L_{V,AV,T}(q a_{_{\rm
    L}})$ are infrared divergent when considered separately, and thus
must be grouped together in order to give finite results. Such groups
are formed by diagrams (3-7), (8-9), (10-11,19) in Fig.
\ref{ZVAT2loopDiagrams} and diagrams (1-2), (3-4) in Fig.
\ref{singletVAT}.  

In Figs. \ref{ZVAT1loopDiagrams} to \ref{singletVAT}, ``mirror''
diagrams (those in which the direction of the external fermion line is
reversed) should also be included. In most cases, these coincide
trivially with the original diagrams; even in the remaining cases,
they can be seen to give equal contribution, by invariance under
charge conjugation. 

As mentioned before, all calculations should be performed at 
vanishing renormalized mass; this can be achieved by working with
massless fermion propagators, provided an appropriate fermion mass
counterterm is introduced (diagram 11 in Fig.
\ref{ZVAT2loopDiagrams}).  

All two-loop diagrams have been calculated in the bare Feynman gauge
($\alpha_\circ=1$). One-loop diagrams have been calculated for generic
values of $\alpha_\circ$; this allows us to convert our two-loop
results to the renormalized Feynman gauge ($\alpha_{RI^{\prime}}=1$ or
$\alpha_{\overline{MS}}=1$). After performing the calculation for the
cases of the vector and axial-vector operator, we see that one-loop
expressions for the renormalization functions do not depend on the
gauge parameter. Especially for the case of the vector operator,
having in mind Eqs. (\ref{gConversion}-\ref{alphaConversion}) and
Eq. (\ref{CV}), this fact causes the lattice results in the $RI'$ and
in the $\overline{MS}$ scheme to coincide.   

Numerical loop integration was carried out by our ``integrator''
program, a {\em metacode} written in Mathematica, for converting
lengthy integrands into efficient Fortran code. Two-loop numerical
integrals were evaluated as sums over finite lattices, of size up to
$L=40$; the results were then extrapolated to
$L\rightarrow\infty$. Extrapolation is the only source of systematic
error; this error can be estimated quite accurately (see,
e.g. Ref. \cite{PST}), given that $L$-dependence of results can only
span a restricted set of functional forms. 

\subsection{One-loop results} 

1-loop results for $Z_\Gamma^{L,RI^{\prime}}$ are presented below in a
generic gauge. As it turns out, only the tensor renormalization
function depends on the gauge parameter, while for all other
operators, one-loop expressions that emerge are gauge independent. The
errors appearing in the expression for $Z_T^{L,RI^{\prime}}$, result
from the $L\to\infty$ extrapolation.       
\begin{eqnarray}
Z_T^{L,RI^{\prime}}= 1 + \frac{g_\circ^2}{16\pi^2}\,c_F &\bigg{[}&-\ln(a_{_{\rm L}}^2 \bar{\mu}^2)+\alpha_\circ -17.018079209(7) \nonumber \\
&& \,\, +3.91333261(4)\,c_{{\rm SW}} + 1.972295300(5)\,c_{{\rm SW}}^2 \bigg{]} \label{ZT1loopRI}
\end{eqnarray}

The corresponding expressions for $Z_V^{L,RI^\prime}$,
$Z_{AV}^{L,RI^\prime}$ can be read off from
Eqs. (\ref{ZV2loopRI}-\ref{ZA2loopRI}) below. In all cases, one-loop
results in the $\overline{MS}$ scheme ($Z_V^{L,\overline{MS}}$,
$Z_{AV}^{L,\overline{MS}}$, $Z_T^{L,\overline{MS}}$) present no
dependence on the gauge parameter. 

\subsection{Two-loop results} 

In order to derive the expressions for the bare Green's functions
$\Sigma^L_{V,AV,T}$, one must evaluate all Feynman diagrams presented
in Figs. \ref{ZVAT1loopDiagrams}-\ref{singletVAT}. The extraction of
$Z_V^{L,Y}$, $Z_{AV}^{L,Y}$ and $Z_T^{L,Y}$ is then straightforward
via Eqs. (\ref{ZXrulesa}-\ref{ZXrulesc}) (for $Y=RI^{\prime}$), and 
via Eqs. (\ref{VTConversion}-\ref{CT}, \ref{AxialConversion}) (for
$Y=\overline{MS}$). To this end, we need the following one-loop 
expression for $Z_A^{L,Y}$ (note that $Z_{\alpha}=1$ to this order):  

\begin{eqnarray}
Z_A^{L,RI^{\prime}}&=& Z_A^{L,\overline{MS}} + {\cal O}\left(g_{\rm o}^4\right) \nonumber \\
&=&1 + \frac{g_\circ^2}{16\pi^2} \bigg[\ln\left(a_{_{\rm L}}^2\bar{\mu}^2\right)\left(\frac{2}{3}\,N_f-\frac{5}{3}\,N_c \right) \nonumber \\
&& \hskip 1.15cm + N_f\,\left(-2.168501047(1) + 0.7969452308(4)\,c_{\rm SW} -4.7126914428(1)\,c_{\rm SW}^2\right) \nonumber \\ 
&& \hskip 1.15cm  + 39.47841760436(1)\,c_F +1.94017130069(1)\,N_c\bigg] + {\cal O}\left(g_{\rm o}^4\right) 
\label{ZA1loop}
\end{eqnarray}

To express our results in terms of the renormalized coupling constant,
we also need the one-loop expression for $Z_g^{L,Y}$:  
\begin{eqnarray}
Z_g^{L,RI^{\prime}}&=& Z_g^{L,\overline{MS}} + {\cal O}\left(g_{\rm o}^4\right) \nonumber \\
&=&1 + \frac{g_\circ^2}{16\pi^2} \bigg[\ln\left(a_{_{\rm L}}^2\bar{\mu}^2\right)\left(-\frac{1}{3}\,N_f+\frac{11}{6}\,N_c \right)  \nonumber \\
&& \hskip 1.15cm + N_f\,\left(0.5286949677(5) -0.3984726154(2)\,c_{\rm SW} + 2.35634572140(7)\,c_{\rm SW}^2\right) \nonumber \\ 
&& \hskip 1.15cm -19.73920880218(1)\,c_F -3.54958342046(1)\,N_c\bigg] + {\cal O}\left(g_{\rm o}^4\right)
\label{Zg1loop}
\end{eqnarray}
Eqs. (\ref{ZA1loop}, \ref{Zg1loop}) are in agreement with older
references (see, e.g., Ref. \cite{Bode}). 

A final necessary ingredient
is the two-loop expression for $Z_{\psi}^{L,RI^{\prime}}$, as required
by Eqs.(\ref{ZXrulesa}-\ref{ZXrulesc}); this was calculated in
Ref. \cite{SP-08}, in the renormalized Feynman 
gauge $\alpha_{RI^{\prime}}=1$, and is included here for completeness:

\newpage

\begin{eqnarray}
Z_{\psi}^{L,RI^{\prime}} = 1 &+& \frac{g_\circ^2}{16\pi^2}\,c_F \bigg{[} \ln(a_{_{\rm L}}^2 \bar{\mu}^2) + 11.852404288(5) - 2.248868528(3)\,c_{{\rm SW}} - 1.397267102(5)\,c_{{\rm SW}}^2 \bigg{]} \nonumber \\
&+& \frac{g_\circ^4}{(16\pi^2)^2}\,c_F \Bigg{[} \ln^2(a_{_{\rm L}}^2 \bar{\mu}^2) \left ( \frac{1}{2} c_F + \frac{2}{3} N_f -\frac{8}{3} N_c \right ) \nonumber \\
&&\quad\quad\,\, + \ln(a_{_{\rm L}}^2 \bar{\mu}^2) \Big{(}-6.36317446(8)\,N_f + 0.79694523(2)\,N_f\,c_{{\rm SW}} \nonumber \\
&& \hskip 3.2cm - 4.712691443(4)\,N_f\,c_{{\rm SW}}^2 \nonumber \\
&& \hskip 3.2cm +49.83082185(5)\,c_F - 2.24886861(7)\,c_F\,c_{{\rm SW}} \nonumber \\
&& \hskip 3.2cm - 1.39726705(1)\,c_F\,c_{{\rm SW}}^2  +29.03029398(4)\,N_c\Big{)} \nonumber \\
&&\quad\quad\,\, + N_f\,\Big{(}-7.838(2) + 1.153(1)\,c_{{\rm SW}} + 3.202(3)\,c_{{\rm SW}}^2 \nonumber \\
&& \hskip 2.2cm +6.2477(6)\,c_{{\rm SW}}^3 + 4.0232(6)\,c_{{\rm SW}}^4 \Big{)}  \nonumber \\
&&\quad\quad\,\, + c_F\,\Big{(}505.39(1) - 58.210(9)\,c_{{\rm SW}} + 20.405(5)\,c_{{\rm SW}}^2 \nonumber \\
&& \hskip 2.2cm +18.8431(8)\,c_{{\rm SW}}^3 + 4.2793(2)\,c_{{\rm SW}}^4 \Big{)}  \nonumber \\
&&\quad\quad\,\, + N_c\,\Big{(}-20.59(1) - 3.190(5)\,c_{{\rm SW}} -
  23.107(6)\,c_{{\rm SW}}^2  \nonumber \\
&& \hskip 2.2cm -5.7234(5)\,c_{{\rm SW}}^3 -
  0.7938(1)\,c_{{\rm SW}}^4 \Big{)} \Bigg{]} 
\label{Zpsi2loopRI}
\end{eqnarray}

We present below $Z_V^{L,RI^{\prime}}$, $Z_{AV}^{L,RI^{\prime}}$ 
and $Z_T^{L,RI^{\prime}}$ to two loops in the renormalized Feynman 
gauge $\alpha_{RI^{\prime}}=1$; we also present the $\overline{MS}$
analogues $Z_{AV}^{L,\overline{MS}}$ and $Z_T^{L,\overline{MS}}$ in the
gauge $\alpha_{\overline{MS}}=1$ (as already mentioned,
$Z_V^{L,\overline{MS}} = Z_V^{L,RI^{\prime}}$). The bare Green's
functions are relegated to Appendix C, where a {\em per diagram} breakdown of the
results is provided. It is a straightforward exercise to recover the
total bare Green's functions from the
corresponding $Z$'s and the renormalized Green's functions.    
\begin{eqnarray}
Z_V^{L,RI^{\prime}} = 1 &+& \frac{g_\circ^2}{16\pi^2}\,c_F
\bigg{[}-20.617798655(6) + 4.745564682(3)\,c_{{\rm SW}} + 0.543168028(5)\,c_{{\rm SW}}^2 \bigg{]} \nonumber \\
&+& \frac{g_\circ^4}{(16\pi^2)^2}\,c_F \bigg{[} N_f\,\Big{(}25.610(3) -11.058(1)\,c_{{\rm SW}} + 33.937(3)\,c_{{\rm SW}}^2 \nonumber \\
&& \hskip 3cm -13.5286(6)\,c_{{\rm SW}}^3 -1.2914(6)\,c_{{\rm SW}}^4 \Big{)}  \nonumber \\
&&\qquad\qquad\quad + c_F\,\Big{(}-539.78(1) -223.57(2)\,c_{{\rm SW}} -104.116(5)\,c_{{\rm SW}}^2  \nonumber \\
&& \hskip 3.2cm -32.2623(8)\,c_{{\rm SW}}^3 +4.5575(3)\,c_{{\rm SW}}^4 \Big{)}  \nonumber \\
&&\qquad\qquad\quad + N_c\,\Big{(}-51.59(1) +18.543(5)\,c_{{\rm SW}} +20.960(6)\,c_{{\rm SW}}^2 \nonumber \\
&& \hskip 3.2cm +2.5121(5)\,c_{{\rm SW}}^3 +0.1765(1)\,c_{{\rm SW}}^4 \Big{)} \bigg{]} 
\label{ZV2loopRI}
\end{eqnarray}

\begin{eqnarray}
Z_{AV}^{L,RI^{\prime}} = 1 &+& \frac{g_\circ^2}{16\pi^2}\,c_F \bigg{[}
  -15.796283066(5) -0.247827627(3)\,c_{{\rm SW}} + 2.251366176(5)\,c_{{\rm SW}}^2 \bigg{]} \nonumber \\
&+& \frac{g_\circ^4}{(16\pi^2)^2}\,c_F \Bigg{[} N_f\,\Big{(}18.497(3) -1.285(1)\,c_{{\rm SW}} + 19.071(3)\,c_{{\rm SW}}^2 \nonumber \\
&& \hskip 3cm +1.0333(6)\,c_{{\rm SW}}^3 -6.7549(6)\,c_{{\rm SW}}^4 \Big{)}  \nonumber \\
&&\hskip 2cm + c_F\,\Big{(}-184.01(1) -389.86(1)\,c_{{\rm SW}} -166.738(6)\,c_{{\rm SW}}^2  \nonumber \\
&& \hskip 3cm +7.894(1)\,c_{{\rm SW}}^3 + 4.3201(3)\,c_{{\rm SW}}^4 \Big{)}  \nonumber \\
&&\hskip 2cm + N_c\,\Big{(}-21.62(1) -33.652(5)\,c_{{\rm SW}} +26.636(6)\,c_{{\rm SW}}^2 \nonumber \\
&& \hskip 3cm +10.2186(5)\,c_{{\rm SW}}^3 +1.4893(1)\,c_{{\rm SW}}^4 \Big{)} \Bigg{]} 
\label{ZA2loopRI}\\[2.0ex]
Z_{AV}^{L,\overline{MS}} = 1 &+& \frac{g_\circ^2}{16\pi^2}\,c_F\,\bigg{[}-11.796283066(5) 
-0.247827627(3)\,c_{{\rm SW}} + 2.251366176(5)\,c_{{\rm SW}}^2 \bigg{]} \nonumber \\
&+& \frac{g_\circ^4}{(16\pi^2)^2}\,c_F\,\Bigg{[}\ln(a_{_{\rm L}}^2 \bar{\mu}^2) 
\left(\frac{8}{3}\,N_f - \frac{44}{3}\,N_c \right) \nonumber \\
&&\hskip 2cm + N_f\,\Big{(}14.045(3) + 1.903(1)\,c_{{\rm SW}}+0.220(3)\,c_{{\rm SW}}^2 \nonumber \\
&& \hskip 3cm +1.0333(6)\,c_{{\rm SW}}^3 -6.7549(6)\,c_{{\rm SW}}^4 \Big{)} \nonumber \\
&&\hskip 2cm + c_F\,\Big(-95.28(1)-390.85(1)\,c_{{\rm SW}} -157.733(6)\,c_{{\rm SW}}^2 \nonumber \\
&& \hskip 3cm +7.894(1)\,c_{{\rm SW}}^3 + 4.3201(3)\,c_{{\rm SW}}^4\Big)   \nonumber \\
&&\hskip 2cm + N_c\,\Big(18.67(1)-33.652(5)\,c_{{\rm SW}}+26.636(6)\,c_{{\rm SW}}^2 \nonumber \\
&& \hskip 3cm +10.2186(5)\,c_{{\rm SW}}^3+1.4893(1)\,c_{{\rm SW}}^4 \Big) \Bigg{]}
\label{ZA2loopMS}
\end{eqnarray}
\begin{eqnarray}
Z_T^{L,RI^{\prime}} = 1 &+& \frac{g_\circ^2}{16\pi^2}\,c_F \bigg{[}
  -\ln(a_{_{\rm L}}^2 \bar{\mu}^2) -16.018079209(7) \nonumber \\[-0.44ex]
&& \hskip 1.7cm +3.91333261(4)\,c_{{\rm SW}} + 1.972295300(5)\,c_{{\rm SW}}^2 \bigg{]} \nonumber \\
&+& \frac{g_\circ^4}{(16\pi^2)^2}\,c_F \Bigg{[} \ln^2(a_{_{\rm L}}^2 \bar{\mu}^2) \left ( \frac{1}{2} c_F -\frac{1}{3} N_f -\frac{11}{6} N_c \right ) \nonumber \\
&&\hskip 2cm + \ln(a_{_{\rm L}}^2 \bar{\mu}^2) \Big{(}3.1685002(6)\,N_f -0.79694524(6)\,N_f\,c_{{\rm SW}} \nonumber \\[-0.44ex]
&& \hskip 4cm +4.71269143(3)\,N_f\,c_{{\rm SW}}^2 \nonumber \\[-0.44ex]
&& \hskip 4cm -13.96033835(7)\,c_F -3.9133325(1)\,c_F\,c_{{\rm SW}} \nonumber \\ [-0.44ex]
&& \hskip 4cm -1.97229535(2)\,c_F\,c_{{\rm SW}}^2 -25.04361149(6)\,N_c\Big{)} \nonumber \\[-0.44ex]
&&\hskip 2cm + N_f\,\Big{(}16.923(6) -8.399(2)\,c_{{\rm SW}} + 18.711(3)\,c_{{\rm SW}}^2 \nonumber \\[-0.44ex]
&& \hskip 3cm -10.8351(8)\,c_{{\rm SW}}^3 -5.1253(6)\,c_{{\rm SW}}^4 \Big{)}  \nonumber \\[-0.44ex]
&&\hskip 2cm + c_F\,\Big{(}-868.0(1) + 551.6(2)\,c_{{\rm SW}} + 63.9(1)\,c_{{\rm SW}}^2  \nonumber \\[-0.44ex]
&& \hskip 3cm -79.49(1)\,c_{{\rm SW}}^3 -12.586(1)\,c_{{\rm SW}}^4 \Big{)}  \nonumber \\[-0.44ex]
&&\hskip 2cm + N_c\,\Big{(}-15.76(8) +27.6(1)\,c_{{\rm SW}} +38.2(1)\,c_{{\rm SW}}^2 \nonumber \\[-0.44ex]
&& \hskip 3cm +7.021(8)\,c_{{\rm SW}}^3 +1.6653(9)\,c_{{\rm SW}}^4 \Big{)} \Bigg{]} 
\label{ZT2loopRI}\\[1.0ex]
Z_T^{L,\overline{MS}} = 1 &+& \frac{g_\circ^2}{16\pi^2}\,c_F\,\bigg{[}
  -\ln(a_{_{\rm L}}^2 \bar{\mu}^2) -17.018079209(7) \nonumber \\[-0.44ex]
&& \hskip 1.7cm +3.91333261(4)\,c_{{\rm SW}} + 1.972295300(5)\,c_{{\rm SW}}^2 \bigg{]} \nonumber \\
&+& \frac{g_\circ^4}{(16\pi^2)^2}\,c_F\,\Bigg{[} \ln^2(a_{_{\rm L}}^2 \bar{\mu}^2) \left(-\frac{1}{3}\,N_f + \frac{1}{2}\,c_F + \frac{11}{6}\,N_c \right)  \nonumber \\
&& \hskip 2cm + \ln(a_{_{\rm L}}^2 \bar{\mu}^2) \Big{(}2.5018336(6)\,N_f -0.79694524(6)\,N_f\,c_{{\rm SW}} \nonumber \\[-0.44ex]
&& \hskip 4cm +4.71269143(3)\,N_f\,c_{{\rm SW}}^2 \nonumber \\[-0.44ex]
&& \hskip 4cm -12.96033835(7)\,c_F -3.9133325(1)\,c_F\,c_{{\rm SW}}\nonumber \\[-0.44ex]
&& \hskip 4cm -1.97229535(2)\,c_F\,c_{{\rm SW}}^2 -21.37694482(6)\,N_c \Big{)} \nonumber \\[-0.44ex]
&&\hskip 2cm + N_f\,\Big{(}21.989(6)-9.196(2)\,c_{{\rm SW}}+23.424(3)\,c_{{\rm SW}}^2 \nonumber \\[-0.44ex]
&& \hskip 3cm -10.8351(8)\,c_{{\rm SW}}^3 -5.1253(6)\,c_{{\rm SW}}^4 \Big{)} \nonumber \\[-0.44ex]
&&\hskip 2cm + c_F\,\Big(-893.2(1)+547.7(2)\,c_{{\rm SW}} +61.9(1)\,c_{{\rm SW}}^2 \nonumber \\[-0.44ex]
&& \hskip 3cm -79.49(1)\,c_{{\rm SW}}^3 -12.586(1)\,c_{{\rm SW}}^4\Big)   \nonumber \\[-0.44ex]
&&\hskip 2cm + N_c\,\Big(-41.44(8)+27.6(1)\,c_{{\rm SW}}+38.2(1)\,c_{{\rm SW}}^2 \nonumber \\[-0.44ex]
&& \hskip 3cm +7.021(8)\,c_{{\rm SW}}^3+1.6653(9)\,c_{{\rm SW}}^4 \Big) \Bigg{]}
\label{ZT2loopMS}
\end{eqnarray}

All expressions reported thus far for $Z_V$, $Z_{AV}$  and $Z_T$ refer
to flavor nonsinglet operators. In the case of $Z_V$ and $Z_T$, all 
diagrams of Fig. \ref{singletVAT} vanish, so that singlet and
nonsinglet results coincide, just as in dimensional
regularization. For 
$Z_{AV}$ on the other hand, the above diagrams give an additional
contribution:  
\begin{eqnarray}
Z_{AV,\,\rm singlet}^{L,RI^{\prime}} = Z_{AV}^{L,RI^{\prime}} 
&+& \frac{g_\circ^4}{(16\pi^2)^2}\,c_F N_f\,
\Bigl( - 6\,\ln(a_{_{\rm L}}^2 \bar{\mu}^2)-2.0491(5) + 15.0315(6)\,c_{{\rm SW}}
\nonumber \\
&& \hskip 1.5cm  
+ 5.0090(2)\,c_{{\rm SW}}^2 - 2.11016(5)\,c_{{\rm SW}}^3 -0.04329(2)\,c_{{\rm SW}}^4\Bigr)
\label{ScalarSingletRI}
\end{eqnarray}
The same extra contribution applies also to the $\overline{MS}$
scheme.  

For the sake of completeness, and as an additional check on our 
results, we compute the renormalized Green's functions (for {\em 
vanishing} renormalized mass). Since the bare Green's functions  
have two contributions of different structure (as defined in
Eq. (\ref{2ptFunct}), see also Eq. (\ref{rngrnfunctn})), we derive the renormalized expressions for 
these contributions separately:    
\begin{eqnarray}
G_V^{(i)\,RI^{\prime}}(q)&\equiv& Z_\psi^{L,RI^{\prime}}\,Z_V^{L,RI^{\prime}}\,\Sigma_V^{(i),L}
\label{Grenormconditionsa} \\
G_{AV}^{(i)\,RI^{\prime}}(q)&\equiv&
Z_\psi^{L,RI^{\prime}}\,Z_{AV}^{L,RI^{\prime}}\,\Sigma_{AV}^{(i),L} 
\label{Grenormconditionsb} \\
G_T^{(i)\,RI^{\prime}}(q)&\equiv&
Z_\psi^{L,RI^{\prime}}\,Z_T^{L,RI^{\prime}}\,\Sigma_T^{(i),L}
\label{Grenormconditionsc} 
\end{eqnarray}
where $i=1,2$. Similarly for $\overline{MS}$, taking into account
Eq. (\ref{RenormGreen}). 

Since these functions are regularization independent, they can be
calculated also using, e.g., dimensional regularization. We have
computed $G_V^{(i)}$, $G_{AV}^{(i)}$ and $G_T^{(i)}$ in both ways: either
starting from our Eqs.(\ref{ZT1loopRI}-\ref{ZT2loopMS}) or using
renormalization functions from dimensional regularization
\cite{Gracey}. In all cases the two ways are in complete agreement. We 
obtain:  
\begin{eqnarray}
G^{(1)\,RI^{\prime}}_V(q)=1 &+& \frac{g_{RI^{\prime}}^2}{16\pi^2}\,c_F\ln(\bar{\mu}^2/q^2) \nonumber \\
&+& \frac{g_{RI^{\prime}}^4}{(16\pi^2)^2}\,c_F\Bigg[\ln^2(\bar{\mu}^2/q^2)\left(\frac{1}{2}\,c_F+N_c\right) \nonumber \\ 
&&\hskip 2cm + \ln(\bar{\mu}^2/q^2)\left(-\frac{19}{9}\,N_f-\frac{3}{2}\,c_F+\frac{251}{18}\,N_c\right)\Bigg]
\label{Vector1RenormGreenFnRI} \\ 
G^{(2)\,RI^{\prime}}_V(q)&=&
\frac{g_{RI^{\prime}}^2}{16\pi^2}\,c_F\left(-2 \ln(\bar{\mu}^2/q^2)\right) \nonumber \\
&+&
\frac{g_{RI^{\prime}}^4}{(16\pi^2)^2}\,c_F\bigg[\ln(\bar{\mu}^2/q^2)\left(-2\,c_F-4\,N_c\right)
+ \frac{38}{9}\,N_f + 3\,c_F-\frac{251}{9}\,N_c\bigg]
\label{Vector2RenormGreenFnRI} 
\end{eqnarray}
The vector renormalized Green's function in the $RI'$ scheme coincides 
with the corresponding axial-vector expression, and thus 
Eqs.(\ref{Vector1RenormGreenFnRI}-\ref{Vector2RenormGreenFnRI}) also 
hold for the axial-vector case: $G^{(1)}_{AV}(q)=G^{(1)}_V(q)$,
$G^{(2)}_{AV}(q)=G^{(2)}_V(q)$. Of course, even though the
$\overline{MS}$ expression for the vector renormalization function,
$Z_V^{L,\overline{MS}}$, coincides with the $RI'$ expression, that is
not the case for the renormalized $\overline{MS}$ Green's function,
due to $C_\psi$ appearing in Eq. (\ref{RenormGreen}). This factor
results in the following quantities:  
\begin{eqnarray}
G^{(1)\,\overline{MS}}_V(q)=1 &+&
\frac{g_{\overline{MS}}^2}{16\pi^2}\,c_F\left(\ln(\bar{\mu}^2/q^2) +1 \right) \nonumber \\
&+& \frac{g_{\overline{MS}}^4}{(16\pi^2)^2}\,c_F\Bigg[\ln^2(\bar{\mu}^2/q^2)\left(\frac{1}{2}\,c_F+N_c\right) \nonumber \\  
&& \hskip 2cm+ \ln(\bar{\mu}^2/q^2)\left(-\frac{19}{9}\,N_f - \frac{1}{2}\,c_F+\frac{251}{18}\,N_c\right) \nonumber \\
&& \hskip 2cm+\left(-\frac{7}{4}\,N_f-\frac{5}{8}\,c_F+\left(\frac{143}{8}-6\zeta(3)\right)\,N_c\right)\Bigg]
\label{Vector1RenormGreenFnMS} \\ 
G^{(2)\,\overline{MS}}_V(q) &=&
-\frac{g_{\overline{MS}}^2}{16\pi^2}\,2\,c_F \nonumber \\
&+&
\frac{g_{\overline{MS}}^4}{(16\pi^2)^2}\,c_F\bigg[\ln(\bar{\mu}^2/q^2)\left(-2\,c_F-4\,N_c\right)
+ \frac{38}{9}\,N_f + c_F-\frac{251}{9}\,N_c\bigg]
\label{Vector2RenormGreenFnMS} 
\end{eqnarray}

Furthermore, the axial-vector renormalized 2-point functions in the
$\overline{MS}$ scheme differ from
Eqs. (\ref{Vector1RenormGreenFnMS}-\ref{Vector2RenormGreenFnMS}), due
to the finite conversion factor $Z_5^{ns}$; they read:  
\begin{eqnarray}
G^{(1)\,\overline{MS}}_{AV}(q)=1 &+&
\frac{g_{\overline{MS}}^2}{16\pi^2}\,c_F\left(\ln(\bar{\mu}^2/q^2) + 5\right) \nonumber \\
&+& \frac{g_{\overline{MS}}^4}{(16\pi^2)^2}\,c_F\Bigg[\ln^2(\bar{\mu}^2/q^2)\left(\frac{1}{2}\,c_F+N_c\right) \nonumber \\  
&& \hskip 2cm+ \ln(\bar{\mu}^2/q^2)\left(-\frac{19}{9}\,N_f + \frac{7}{2}\,c_F+\frac{251}{18}\,N_c\right) \nonumber \\
&& \hskip 2cm+\left(-\frac{71}{36}\,N_f-\frac{21}{8}\,c_F+\left(\frac{2143}{72}-6\zeta(3)\right)\,N_c\right)\Bigg]
\label{Axial1RenormGreenFnMS} \\ 
G^{(2)\,\overline{MS}}_{AV}(q)&=&
-\frac{g_{\overline{MS}}^2}{16\pi^2}\,2\,c_F \nonumber \\
&+&
\frac{g_{\overline{MS}}^4}{(16\pi^2)^2}\,c_F\bigg[\ln(\bar{\mu}^2/q^2)\left(-2\,c_F-4\,N_c\right)
+ \frac{38}{9}\,N_f - 7\,c_F-\frac{251}{9}\,N_c\bigg]
\label{Axial2RenormGreenFnMS} 
\end{eqnarray}
If one considers the singlet axial-vector current, then there exists
an extra contribution to the expressions above:
\begin{eqnarray}
G^{(1),\overline{MS}}_{AV,\,\rm singlet}(q) &=& G^{(1)\,\overline{MS}}_{AV}(q)
+ \frac{g_{\overline{MS}}^4}{(16\pi^2)^2}\,
\left( - 6\,\ln(\bar{\mu}^2/q^2)\,c_F N_f - \frac{3}{2}\,c_F N_f\,\right)
\label{Axial1SingletRenormMS} \\
G^{(2),\overline{MS}}_{AV,\,\rm singlet}(q) &=& G^{(2)\,\overline{MS}}_{AV}(q)
+\frac{g_{\overline{MS}}^4}{(16\pi^2)^2}\,\left(-4\,c_F N_f\right)
\label{Axial2SingletRenormMS}
\end{eqnarray}
For the $RI'$ scheme, similar relations hold, the only difference
being the absence of the factors $Z_5^s$, $Z_5^{ns}$; we obtain:
\begin{eqnarray}
G^{(1),\,RI^{\prime}}_{AV,\,\rm singlet}(q) &=& G^{(1)\,RI^{\prime}}_{AV}(q)
+ \frac{g_{RI^{\prime}}^4}{(16\pi^2)^2}\,
\left( - 6\,\ln(\bar{\mu}^2/q^2)\,c_F N_f\,\right)
\label{Axial1SingletRenormRI} \\
G^{(2),\,RI^{\prime}}_{AV,\,\rm singlet}(q) &=& G^{(2)\,RI^{\prime}}_{AV}(q)
+\frac{g_{RI^{\prime}}^4}{(16\pi^2)^2}\,\left(-4\,c_F N_f\right)
\label{Axial2SingletRenormRI}
\end{eqnarray}
Finally, for the tensor renormalized Green's function, we obtain:  
\begin{eqnarray}
G^{(1)\,RI^{\prime}}_T(q)=1 &+& \frac{g_{RI^{\prime}}^4}{(16\pi^2)^2}\,c_F\Bigg[\ln^2(\bar{\mu}^2/q^2)\left(\frac{1}{3}\,N_f - \frac{5}{6}\,N_c\right) \nonumber \\
&&\hskip 2cm + \ln(\bar{\mu}^2/q^2)\left(-\frac{2}{3}\,N_f + 8\,c_F - \frac{7}{3}\,N_c \right)\Bigg]
\label{Tensor1RenormGreenFnRI}
\end{eqnarray}
Just as was expected from dimensional regularization, $G^{(2)\,RI^{\prime}}_T(q)=0$. The
corresponding quantity in the $\overline{MS}$ scheme reads: 
\begin{eqnarray}
G^{(1)\,\overline{MS}}_T(q)=1 &+& \frac{g_{\overline{MS}}^4}{(16\pi^2)^2}\,c_F\Bigg[\ln^2(\bar{\mu}^2/q^2)\left(\frac{1}{3}\,N_f - \frac{5}{6}\,N_c\right) \nonumber \\
&&\hskip 2cm + \ln(\bar{\mu}^2/q^2)\left(-\frac{2}{3}\,N_f + 8\,c_F - \frac{7}{3}\,N_c \right) \nonumber \\
&&\hskip 2cm + \frac{31}{27}\,N_f +\left(\frac{62}{3}-20\,\zeta(3)\right)\,c_F + \left(-\frac{761}{54} + 8\,\zeta(3)\right)\Bigg]
\label{Tensor1RenormGreenFnMS} 
\end{eqnarray}

In Figs. \ref{ZVRIplot}, (\ref{ZARIplot},\ref{ZAMSplot}),
(\ref{ZTRIplot},\ref{ZTMSplot}) we plot $Z_V^{L,RI^{\prime}}$,
($Z_{AV}^{L,RI^{\prime}}$, $Z_{AV}^{L,\overline{MS}}$) and
($Z_T^{L,RI^{\prime}}$, $Z_T^{L,\overline{MS}}$), respectively, as a
function of $c_{\rm SW}$. Values of the clover parameter used in
simulations lie within the typical range $0\le 
c_{\rm SW} \lesssim 2$. For definiteness, we have set $N_c=3$,
$\bar{\mu}=1/a_{_{\rm L}}$ and $\beta_{\rm o}\equiv 2N_c/g_{\rm
  o}^2=6.0$. Our results up to two loops for each $Z$ are shown for
both $N_f=0$ and $N_f=2$, and compared to the corresponding one-loop
results. Furthermore, in the axial-vector case, we also present the
two-loop result for the flavor singlet operator, for $N_f=2$.

In Fig. \ref{ZVATRIplot} we present, on the same plot, the values of   
$Z_V^{L,RI^{\prime}}$, $Z_{AV}^{L,RI^{\prime}}$, 
$Z_{AV,\,singlet}^{L,RI^{\prime}}$ and $Z_T^{L,RI^{\prime}}$ up to 2
loops, versus $c_{\rm SW}$. We have chosen $N_c=3$,
$\bar{\mu}=1/a_{_{\rm L}}$, $N_f=2$ and $\beta_{\rm o}=5.3$. The
corresponding results in the $\overline{MS}$ scheme are plotted in
Fig. \ref{ZVATMSplot}.

\newpage

\section{Discussion}
\label{Discussion}

In this paper we have reported results regarding the
Vector, Axial-Vector and Tensor fermion bilinear operators. This work,
along with a previously published paper \cite{SP-08} regarding the
Scalar and Pseudoscalar operators, provide a complete two loop
calculation for the renormalization functions for local fermion
bilinears, considering both the singlet and nonsinglet cases. The
two-loop wave function renormalization constant, $Z_\psi$, which is a
prerequisite for our calculation, was presented in Ref. \cite{SP-08}
(the reader should also refer to this paper for any necessary notation
not included in the present sequel paper). 

It is clear from Figs. \ref{ZVRIplot} to \ref{ZTMSplot} that the
two-loop renormalization functions differ significantly from 1-loop
values; this difference must then be properly taken into account in
reducing systematic error in MC simulations. At the same time, 2-loop
contributions are typically smaller than 1-loop contributions,
especially for $c_{SW} \lesssim 1$, indicating that the (asymptotic)
perturbative series are under control.  

The results are presented as a function of the clover parameter, where
the values of $c_{SW}$ lie within the standard range $0\leq c_{SW} \leq 2$. 
Optimal values for $c_{SW}$, which have been estimated both
non-perturbatively \cite{Luscher1996} and perturbatively (to 1-loop)
\cite{SW}, lie within this range. A breakdown of our results on a {\em
per diagram} basis is presented in Appendix C, for completeness.

As already mentioned, we take into account both singlet and nonsinglet
operators. After evaluating all Feynman diagrams involved, we found
that, for the Vector and Tensor operators, singlet renormalization
functions coincide with nonsinglet ones. On the other hand, the
Axial-Vector operator receives an additional contribution in the
flavor singlet case.

The numerical integrations over loop momenta were executed on a
Pentium IV cluster; they required the equivalent of 60 months on a
single CPU.

A possible extension to the present calculation is the renormalization
of more extended operators, with the same continuum limit as we have
considered here. A standard basis of higher dimension operators, with
the same quantum numbers as the local bilinears which we have
considered, can be found e.g. in Ref. \cite{Capitani}. Such operators
are frequently used to reduce ${\cal O}(a_{_{\rm L}})$ effects. A
number of additional Feynman diagrams must be introduced, since the
vertices coming from these operators may also contain gluon
lines. However, the
additional integrals resulting after the contractions will be free of
superficial divergences, leading to a less cumbersome computation,
despite an increase in the size of the integrals. Further directions
regard higher dimensional operators, such as $\bar{\Psi}\,D^\mu\cdots 
D^\nu\,\Gamma\Psi$, which enter structure function calculations, and
4-fermion operators. 

Finally, our computation can be easily extended to improved lattice
actions. With regard to improved fermion actions, such as those
containing twisted mass terms \cite{Frezzotti} or \"Osterwalder-Seiler
terms \cite{Rossi}, our results remain unchanged, since they pertain
to mass-independent schemes. Improving the gluon action, on the other 
hand, is more CPU consuming, but conceptually straightforward:
Splitting (in iterative fashion) the Symanzik propagator into a Wilson 
gluon propagator plus the remainder, leads to the same bare Green's
functions as the ones presented in this paper, with the addition of
superficially convergent terms, which can be more easily manipulated.
Based on our experience with other similar calculations, the algebraic
expressions for the integrands will grow roughly by a factor of 5; furthermore,
the gluon propagator must now be inverted numerically for each
value of the momentum, leading to an additional factor of $\lesssim$2 in
CPU time. Finally, if one wishes to employ more than one set of values
for the Symanzik coefficients, CPU time for numerical integration will
increase almost proportionately.

\newpage
\appendix
\section{Fermions in an Arbitrary Representation}

The results presented up to this point regarded renormalization
constants of various fermion bilinear operators constructed with
fermions in the fundamental representation of the gauge group. Our
results were expressed in terms of the clover parameter, $c_{\rm SW}$,
the number of fermions, $N_f$, the number of colors, $N_c$, and the
quadratic Casimir operator in the fundamental representation, $c_F$.  

Recently there has been interest in theories with fermions in other
representations; some preliminary non-perturbative calculations have
also appeared (see e.g. \cite{Catterall}, \cite{DelDebbio}). In this
Appendix we describe the conventions we use in our work, regarding the
generators of the algebra, and we then express our findings in an
arbitrary representation.     

Our results for $Z_V$, $Z_{AV}$, $Z_T$, Eqs.(\ref{ZV2loopRI},
\ref{ZA2loopRI}, \ref{ZT2loopRI}), can be easily generalized to an
action with Wilson/clover fermions in an arbitrary representation $R$,
of dimensionality $d_R$. 

In this case, the gluon part of the action remains the same, while all 
link variables appearing in the fermion part of the action assume the 
form:
\begin{equation}
U_{x,\,x+\mu} = {\rm exp}(i\,g_0\,A^a_\mu(x)\,T^a)\quad\longrightarrow\quad
U_{x,\,x+\mu} = {\rm exp}(i\,g_0\,A^a_\mu(x)\,T^a_R)
\end{equation}
Using standard notation and conventions, the generators $T^a$ in
the fundamental representation satisfy:
\begin{equation}
[T^a,T^b] = i\,f^{abc}\,T^c,\quad \sum_aT^aT^a 
\equiv \openone\cdot c_F = \openone\cdot {N_c^2-1\over 2N_c}\,,
\quad {\rm tr}(T^aT^b)\equiv\delta^{ab}\,t_F=
\delta^{ab}\,{1\over2}
\end{equation}
In the representation $R$ we have:
\begin{equation}
[T^a_R,T^b_R] = i\,f^{abc}\,T^c_R,\quad \sum_aT^a_RT^a_R 
\equiv \openone\cdot c_R,\quad
{\rm tr}(T^a_RT^b_R)\equiv\delta^{ab}\,t_R
\end{equation}
where: $t_R=(d_R\,c_R)/(N_c^2-1)$.

For the 1-loop quantities, Eqs. (\ref{ZA1loop}, \ref{Zg1loop}),
converting to the representation $R$ is a straightforward
substitution:
\begin{equation}
N_f \longrightarrow N_f \cdot (t_R/t_F)=N_f \cdot (2\,t_R)
\label{tR}
\end{equation}
and, in addition, for Eq. (\ref{ZT1loopRI}):
\begin{equation}
c_F \longrightarrow c_R
\label{cR}
\end{equation}
Aside from these changes, all algebraic expressions (and the numerical
coefficients resulting from loop integrations) remain the same.

A similar reasoning applies to the 2-loop quantities in
Eqs. (\ref{ZV2loopRI}, \ref{ZA2loopRI}, \ref{ZT2loopRI}): For most
diagrams, once their value is expressed as a linear combination of
$c_F^2$, $c_F N_c$ and $c_F N_f$, it suffices to apply substitutions
(\ref{tR}) and (\ref{cR}). The only exceptions are diagrams containing
a gluon tadpole [diagram 3 of Fig. \ref{ZVAT2loopDiagrams}; 1-loop
diagrams, when expressed in terms of $a_{RI^{\prime}}$,
$\alpha_{RI^{\prime}}$ by means of $Z_g$, $Z_A$]: In these cases, only
one power of $c_F$ should be changed to $c_R$; a possible additional
power of $c_F$ originates from the gluon tadpole and should stay as
is. This peculiarity implies that, in order to perform the
substitutions as described above, one must start from the {\em per
diagram} breakdown of 2-loop results. To avoid a lengthy presentation,
we apply, instead, substitutions (\ref{tR}) and (\ref{cR}) 
indiscriminately on Eqs. (\ref{ZV2loopRI}, \ref{ZA2loopRI},
\ref{ZT2loopRI}); consequently, we must then add a correction term, as
follows: 
\begin{eqnarray}
Z_V^{L,RI^{\prime}}\big|_R &=&
Z_V^{L,RI^{\prime}}\big|_{c_F\to c_R\,,\ N_f\to 2 N_f\,t_R} \nonumber \\
&&+\frac{g_\circ^4}{(16\pi^2)^2}\,c_R\,(c_R-c_F) \cdot
\big[813.9580654(2) -187.3473843(1)\,c_{{\rm SW}}\nonumber \\[-0.5ex]
&&\hskip 4.5cm -21.4434142(2)\,c_{{\rm SW}}^2 \big] \\
Z_{AV}^{L,RI^{\prime}}\big|_R &=&
Z_{AV}^{L,RI^{\prime}}\big|_{c_F\to c_R\,,\ N_f\to 2 N_f\,t_R} \nonumber \\
&&+\frac{g_\circ^4}{(16\pi^2)^2}\,c_R\,(c_R-c_F) \cdot
\big[623.6122595(2) + 9.7838425(1)\,c_{{\rm SW}}\nonumber \\[-0.5ex]
&&\hskip 4.5cm -88.8803741(2)\,c_{{\rm SW}}^2 \big] \\
Z_T^{L,RI^{\prime}}\big|_R &=&
Z_T^{L,RI^{\prime}}\big|_{c_F\to c_R\,,\ N_f\to 2 N_f\,t_R}
\nonumber\\
&&+\frac{g_\circ^4}{(16\pi^2)^2}\,c_R\,(c_R-c_F) \cdot \big[ 4 \pi^2 \ln(a_{_{\rm L}}^2 \bar{\mu}^2)
  +632.3684202(3)\nonumber\\[-0.5ex]
&&\hskip 4.5cm -154.492179(2)\,c_{{\rm SW}} 
  -77.8630975(2)\,c_{{\rm SW}}^2 \big]
\end{eqnarray}

Actually, the reader could arrive at these results without knowledge 
of the per diagram breakdown, by virtue of the following fact: All
``exceptional'' powers of $c_F$ cancel out of $Z_V^{L,RI^{\prime}}$,
$Z_{AV}^{L,RI^{\prime}}$, $Z_T^{L,RI^{\prime}}$, if these are expressed
in terms of the renormalized coupling constant
$a_{RI^{\prime}}$. Thus, one may:   
\begin{itemize}
\item[$\bullet$] Express Eqs. (\ref{ZV2loopRI}, \ref{ZA2loopRI},
\ref{ZT2loopRI}) in terms of $g_{RI^{\prime}}$ by means of $g_{\rm
o} =(Z_g^{L,RI^{\prime}})\,g_{RI^{\prime}}$, with
$Z_g^{L,RI^{\prime}}$ in the fundamental representation
(Eq. (\ref{Zg1loop}))  
\item[$\bullet$] Apply substitutions (\ref{tR}), (\ref{cR}) throughout
\item[$\bullet$] If desired, reexpress everything in terms of $g_{\rm o}$ 
(using $(Z_g^{L,RI^{\prime}})^{-1}$ from Eq. (\ref{Zg1loop}), with
$N_f\to 2N_ft_R$ and $c_F$ as is)
\end{itemize}
No correction terms are necessary in this procedure.

\section{Manipulation of superficially divergent and subdivergent terms}

In the case of the Scalar and Pseudoscalar bilinears \cite{SP-08} all
superficially divergent terms involved at most one free Lorentz index,
but when one considers other bilinear operators different structures
may arise. For example, in the case of the vector and axial-vector
operators, two-index integrals may arise and, of course, when working
with the tensor operator three- or even four-index integrals appear
during the manipulation of superficially divergent terms. The most
laborious aspect of such an evaluation is to extract the explicit
dependence of the bare matrix element on the external momentum, by
expressing the superficially divergent parts in terms of known
primitive divergent integrals.  

We will focus on an arbitrary four-index integrand emerging, for
example, from a ``diamond''-like diagram with a $\Gamma =
\gamma_5\,\sigma_{\mu\,\nu}$ insertion. Taking into account the
symmetries of this object, we can deduce all possible tensor
structures which may appear, as a linear combination, in the result
for the corresponding integral. Such structures are certainly tensors
under the hypercubic group, but not necessarily so under the full
$SO(4)$ Euclidean rotation group: Terms such as
$\delta^{\mu\,\nu\,\rho\,\sigma}$ or $q^4/(q^2)^2$ might be
present\footnote{$\delta^{\mu\,\nu\,\rho\,\sigma}\equiv 1$,
  $\mu=\nu=\rho=\sigma$; $\delta^{\mu\,\nu\,\rho\,\sigma}=0$,
otherwise; $q^4\equiv \sum_\mu(q^\mu)^4$.} ($q$: external momentum), and
if so they might spoil the renormalizability and/or the Lorentz
invariance of the theory. We must show that in all cases, such terms
are absent. 

Let us begin by taking as an example an algebraic expression which
contains both superficial and sub divergencies; this example serves as
a prototype for all the cases we have encountered. Such an expression
may arise from a ``diamond''-like diagram with the insertion
 $\Gamma = \gamma_5\,\sigma_{\mu\,\nu}$: 
\begin{equation}
{\phantom{a} \atop \phantom{a}^{\psfig{figure=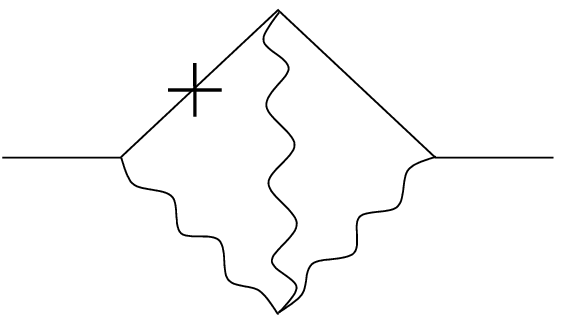, scale=0.35}}} \qquad \longrightarrow \qquad I^{\mu\,\nu\,\rho\,\sigma}(q) = \int\frac{d^4p\,d^4k}{(2 \pi)^8}\,
\frac{\kcirc^{\mu}\,\pcirc^{\nu}\pcirc^{\rho}\,\kcirc^{\sigma}}
{(\hat{p}^2)^2 \ \widehat{p+q}^2 \ \widehat{k-p}^2 \ \hat{k}^2 \ \widehat{k+q}^2}
\label{fourindxobj}
\end{equation}
where $q$ is the external momentum and
\begin{equation}
\hat{p}^\mu=2\sin (\frac{p^\mu}{2})\qquad, \qquad\hat{p}^2=\sum_\mu 4 \sin^2(\frac{p^{\,\mu}}{2})\qquad , \qquad
\pcirc^\mu=\sin(p^{\,\mu})
\label{notation4indx}
\end{equation}
No summation is implied over the indices $\mu$, $\nu$,
$\rho$, $\sigma$. 

From simple ultraviolet power counting on the term above, one can
realize that the superficial degree of divergence is $-8$ and the
degree of divergence in each of the two loops is $-6$ and $-4$. Thus,
this term is not only superficially divergent but also contains a
subdivergence in the right loop.  All divergences are resolved by
using a BPHZ procedure. The potential IR divergences, which may arise
in intermediate steps, necessitate working in $D>4$ dimensions as in
\cite{KNS}. Performing a BPHZ subtraction for the right loop, we split
the integral into two parts: 
\begin{eqnarray}
I^{\mu\,\nu\,\rho\,\sigma}(q)&=&I_{\rm sub}^{\mu\,\nu\,\rho\,\sigma}(q) + \int \frac{d^D p}{(2 \pi)^D}
\frac{\pcirc^{\nu}\,\pcirc^{\rho}}{(\hat{p}^2)^2\ \widehat{p+q}^2}
\int \frac{d^D k}{(2\pi)^D}\frac{\kcirc^{\mu}\,\kcirc^{\sigma}}{(\hat{k}^2)^3} 
\label{FirstSubA} \\[2.0ex]
I_{\rm sub}^{\mu\,\nu\,\rho\,\sigma}(q)&\equiv& \Bigg[\int \frac{d^D p\,d^D k}{(2 \pi)^{2D}} 
\frac{\kcirc^{\mu}\,\kcirc^{\sigma}\,\pcirc^{\nu}\,\pcirc^{\rho}}{(\hat{p}^2)^2\ \widehat{p+q}^2\ \widehat{k-p}^2\ \hat{k}^2\ \widehat{k+q}^2} \nonumber \\
&& \hskip 1.5cm - \int \frac{d^D p}{(2 \pi)^D}\frac{\pcirc^{\nu}\,\pcirc^{\rho}}{(\hat{p}^2)^2\ \widehat{p+q}^2}
\int \frac{d^D k}{(2\pi)^D}\frac{\kcirc^{\mu}\,\kcirc^{\sigma}}{(\hat{k}^2)^3} 
\Bigg ] 
\label{FirstSubB}
\end{eqnarray}
The last term in Eq. (\ref{FirstSubA}) is a separable integral. The
integral over momentum $p$ is a standard primitively divergent
integral (see, e.g., \cite{Luscher}), whose value contains only
Lorentz invariant structures. The integral over momentum $k$ does not
depend on the external momentum $q$, and gives nonzero result only
when the indices $\mu$ and $\sigma$ are in the same direction. Thus,
this term assumes the following structural form: 
\begin{equation}
\int \frac{d^D p}{(2
  \pi)^D}\frac{\pcirc^{\nu}\,\pcirc^{\rho}}{(\hat{p}^2)^2\ \widehat{p+q}^2}
\int \frac{d^D k}{(2\pi)^D}\frac{\kcirc^{\mu}\,\kcirc^{\sigma}}{(\hat{k}^2)^3}
\qquad \longrightarrow \qquad
  \delta^{\mu\,\sigma}\,\left(a\,\delta^{\nu\,\rho}+b\,\frac{q^\nu
  \,q^\rho}{q^2}\right) 
\label{separint1}
\end{equation}
In the remaining part of the original expression, we must still
perform an extra subtraction, to cure the superficial divergence:
\begin{equation}
I_{\rm sub}(q) = \left[ I_{\rm sub}(q) - I_{\rm sub}(0) \right] + I_{\rm sub}(0)
\label{SecondSub}
\end{equation}
According to the BPHZ procedure, the quantity $\left[I_{\rm
sub}(q)-I_{\rm sub}(0)\right]$ is now UV-finite, and thus it equals
the corresponding continuum expression. Consequently, once again, only
Lorentz invariant structures arise:
\begin{eqnarray}
\left[I_{\rm sub}(q)-I_{\rm sub}(0)\right] &\longrightarrow&
a^\prime\,\frac{q^\mu\,q^\nu\,q^\rho\,q^\sigma}{(q^2)^2} +
\frac{b^\prime}{q^2}\,\left(\delta^{\mu\,\nu}\,q^\rho\,q^\sigma +
\delta^{\sigma\,\nu}\,q^\rho\,q^\mu +
\delta^{\mu\,\rho}\,q^\nu\,q^\sigma +
\delta^{\sigma\,\rho}\,q^\mu\,q^\nu \right) \nonumber \\
&& \hskip - 1cm + c^\prime\,\delta^{\mu\,\sigma}\,\frac{q^\nu\,q^\rho}{q^2} +
d^\prime\,\delta^{\nu\,\rho}\,\frac{q^\mu\,q^\sigma}{q^2} +
e^\prime\,\left(\delta^{\mu\,\nu}\,\delta^{\rho\,\sigma} +
\delta^{\sigma\,\nu}\,\delta^{\mu\,\rho}\right) + f^\prime\,\delta^{\mu\,\sigma}\,\delta^{\nu\,\rho}
\label{contintegr}
\end{eqnarray}

The last part of the integral $I^{\mu\,\nu\,\rho\,\sigma}(q)$ (last
term in Eq. (\ref{SecondSub})) equals:  
\begin{equation}
I_{\rm sub}^{\mu\,\nu\,\rho\,\sigma}(0) = 
\int\frac{d^Dp\,d^Dk}{(2\pi)^(2D)}\,\frac{\kcirc^\mu\,\pcirc^\nu\,\pcirc^\rho\,\kcirc^\sigma}
{(\hat{p}^2)^3\ (\hat{k}^2)^2}\, \left(\frac{1}{\widehat{k-p}^2} -
\frac{1}{\hat{k}^2}\right)     
\label{lastpart}
\end{equation}
This $q$-independent integral could give rise to a structural form of
the type $\delta^{\mu\,\nu\,\rho\,\sigma}$, which would spoil Lorentz
invariance; however, this problem is avoided since the indices
$\mu,\,\nu$ in $I^{\mu\,\nu\,\rho\,\sigma}$ actually originate from
the insertion $\Gamma=\gamma_5\,\sigma_{\mu\,\nu}$. As a consequence,
only the combination
$I^{\mu\,\nu\,\rho\,\sigma}-I^{\nu\,\mu\,\rho\,\sigma}$ appears in the
Feynman diagram, and no $\delta^{\mu\,\nu\,\rho\,\sigma}$ contribution
survives. Thus, we are led to:
\begin{equation}
\left[I_{\rm sub}^{\mu\,\nu\,\rho\,\sigma}(0) 
- I_{\rm sub}^{\nu\,\mu\,\rho\,\sigma}(0)\right] \longrightarrow
a\,\left(\delta^{\mu\,\rho}\,\delta^{\nu\,\sigma} 
- \delta^{\nu\,\rho}\,\delta^{\mu\,\sigma} \right)
\label{lastpart2}
\end{equation}
Having completed the whole procedure for the integral shown in
Eq. (\ref{fourindxobj}), we conclude that the only functional form
that a four-index object (with the symmetries described above) can have
reads: 
\begin{eqnarray}
I^{\mu\,\nu\,\rho\,\sigma}(q)- I^{\nu\,\mu\,\rho\,\sigma}(q)&=& 
A\left(q^2\right)\,\left(\delta^{\mu\,\rho}\,\delta^{\nu\,\sigma} 
- \delta^{\nu\,\rho}\,\delta^{\mu\,\sigma} \right) \nonumber \\[2.0ex]
&+&\frac{B\left(q^2\right)}{q^2}\,\Big(\delta^{\mu\,\rho}\,q^\nu\,q^\sigma
-\delta^{\nu\,\rho}\,q^\mu\,q^\sigma \Big) \nonumber \\[2.0ex]
&+& \frac{C\left(q^2\right)}{q^2}\,\Big(\delta^{\mu\,\sigma}\,q^\nu\,q^\rho
-\delta^{\nu\,\sigma}\,q^\mu\,q^\rho\Big)
\label{structform4indx}
\end{eqnarray}
We emphasize again that, even though the above expression would be
obvious in a continuum regularization, it is not so on the lattice,
where one could have ended up with terms breaking Lorentz invariance.

Using similar considerations, we can prove that the two- and
three-index expressions which appear in our calculation, will take the same
structural form as in the continuum; i.e., they will be free of
Lorentz non invariant contributions, which could be present a priori,
such as $\delta^{\mu\,\nu}\,(q^\mu)^2/q^2$, $\delta^{\mu\,\nu\,\rho}$, etc.

Once we establish the structural form of the two-, three- and
four-index integrals, we must compute the coefficients multiplying
each tensor structure, such as the coefficients $A(q^2)$, $B(q^2)$,
$C(q^2)$ of Eq. (\ref{structform4indx}). We illustrate the procedure
by taking as an example the following two-index integral:
\begin{equation}
I^{\mu \,\nu}(q)= \int \frac{d^4 p\,d^4 k}{(2 \pi)^8} 
\frac{\kcirc^{\mu}\,\pcirc^{\nu}\,(\kcirc\cdot\pcirc)}{(\hat{p}^2)^2\
  \widehat{p+q}^2\ \widehat{k-p}^2\ \hat{k}^2\ \widehat{k+q}^2}
\label{TwoIndexTerm}
\end{equation}

Along the same lines of reasoning as above, we conclude that
the lattice integral $I^{\mu\,\nu}$ is of the same form as its
continuum counterpart:
\begin{equation}
I^{\mu \,\nu}(q) = A\,\delta^{\mu\,\nu} + B\,\frac{q_\mu q_\nu}{q^2} 
\label{TwoIndexTerm2}
\end{equation}
The problem is now reduced to evaluating the coefficients $A$ and
$B$. Upon contracting the integral shown in Eq. (\ref{TwoIndexTerm2})
with $\delta^{\mu\,\nu}$ and $q_\mu q_\nu$, we get:
\begin{eqnarray}
I_1&\equiv&\sum_{\mu\,\nu}\delta^{\mu\,\nu}\,I^{\mu \,\nu}=D\,A + B \label{contract1} \\
I_2&\equiv&\sum_{\mu\,\nu}q_\mu\,q_\nu\,I^{\mu \,\nu} = A\,q^2 + B\,q^2 
\label{contract2}
\end{eqnarray}
where $D$ is the number of dimensions (on the lattice, $D=4$). Once we 
evaluate the integrals $I_1$ and $I_2$, we are able to determine the
quantities $A$ and $B$ through the following relations:  
\begin{equation}
A = \frac{1}{3} \left(I_1 - \frac{1}{q^2}\,I_2\right) \qquad , 
\qquad B = \frac{1}{3} \left(\frac{4}{q^2}\,I_2 - I_1\right)
\label{ABvalues}
\end{equation} 

Let us proceed with the evaluation of the integral $I_1$ by
contracting $\delta^{\mu\,\nu}$ with Eq. (\ref{TwoIndexTerm}),
keeping only terms of order ${\cal O}(q^0)$. At first, we aim to
reduce the number of propagators appearing in the denominator by
employing the following property, which is valid on the lattice:
\begin{equation}
\kcirc\cdot\pcirc = \frac{1}{2}\,\left[\widehat{k}^2 + \widehat{p}^2 -
(\widehat{k-p})^2 - \frac{1}{2}\sum_\rho (\hat{k}^\rho)^2\,(\hat{p}^\rho)^2\right] 
\label{dotprod}
\end{equation}
Omitting some intermediate steps, the resulting expression can be
written as follows: 
\begin{equation}
I_1 = I_a + I_b + I_c + I_d
\end{equation}
where
\begin{eqnarray}
I_a &=& -\frac{1}{4}\int \frac{d^4 p\,d^4 k}{(2
  \pi)^8}\,\frac{(\kcirc\cdot\pcirc)\,\sum_\rho (\hat{k}^\rho)^2\,(\hat{p}^\rho)^2}{(\hat{p}^2)^2\
  \widehat{p+q}^2\ \widehat{k-p}^2\ \hat{k}^2\ \widehat{k+q}^2}
  \label{Ia} \\[2.0ex]
I_b &=&-\frac{1}{2}\sum_\rho
\int\frac{d^4k}{(2\pi)^4}\,\frac{\kcirc^\rho}{\hat{k}^2\
  \widehat{k+q}^2}\cdot \int\frac{d^4p}{(2\pi)^4}\,\frac{\pcirc^\rho}{(\hat{p}^2)^2\
  \widehat{p+q}^2} \label{Ib} \\[2.0ex]
I_c &=& \frac{1}{2} 
\int \frac{d^4 p\,d^4 k}{(2 \pi)^8}\,\frac{\kcirc\cdot\pcirc}{\hat{k}^2\
  \widehat{k+q}^2\ \hat{p}^2\ \widehat{p+q}^2\ \widehat{k-p}^2} \ \ \label{Ic}\\[2.0ex]
I_d &=& \int \frac{d^4 p\,d^4
  k}{(2\pi)^8}\,\frac{\kcirc\cdot\pcirc}{(\hat{p}^2)^2\ \widehat{p+q}^2\ \widehat{k+q}^2\ \widehat{k-p}^2} =
\frac{1}{2}\int \frac{d^4 p\,d^4 k}{(2\pi)^8}
\,\frac{(k\, \minuscirc\, q)\cdot(p\, \minuscirc\, q)}{\hat{p}^2\
  \Big(\widehat{p-q}^2\Big)^2\ \hat{k}^2\ \widehat{k-p}^2}\ \ \
\label{Id}
\end{eqnarray}

As can be seen from the expressions above, we have managed to reduce
diamond-like expressions into simpler integrals. Integral $I_a$ is IR
convergent: one can set $q=0$ and carry out the integration
numerically. Integral $I_b$, shown in Eq. (\ref{Ib}), is the product
of two 1-loop integrals: The first one is well known and tabulated in
\cite{Luscher}, whereas the second, being UV-finite, equals its
continuum analogue and can be solved by using the following formula
found in Ref. \cite{Chetyrkin}:   
\begin{equation}
\int \frac{d^D p}{(2 \pi)^D}\frac{{\cal P}_n(p)}{p^{2\alpha}\,(p-q)^{2\beta}}
= \frac{(q^2)^{2-\varepsilon-\alpha-\beta}}{(4 \pi^2)}\,\sum_{\sigma
  \geq 0} G(\alpha,\,\beta,\,n,\,\sigma)\,q^{2\sigma}\,\left[
\frac{1}{\sigma !}
\left(\frac{\square_p}{4}\right)^{\sigma}\,{\cal P}_n(p)\right]_{p=q} 
\label{ChetyrkinInt} 
\end{equation}
where $\varepsilon=(4-D)/2$, $\square_p\equiv\partial^2/\partial
p_\mu\partial p_\mu$, and 
$$
G(\alpha,\,\beta,\,n,\,\sigma)=(4 \pi)^\varepsilon\,\frac{\Gamma(\alpha+\beta-\sigma-2+\varepsilon)}{\Gamma(\alpha)\,\Gamma(\beta)}\,B(2-\varepsilon-\alpha+n-\sigma,\,2-\varepsilon-\beta+\sigma)  
$$
$\Gamma(a)$ is the Gamma function and
$B(\alpha,\beta)=\Gamma(\alpha)\,\Gamma(\beta)/\Gamma(\alpha+\beta)$
is the Beta function.
${\cal P}_n(p)$ is an arbitrary homogeneous polynomial in $p$:
${\cal P}_n(\lambda\,p)=\lambda^n{\cal P}_n(p)$. All UV-convergent
integrals in our calculation can be treated using
Eq. (\ref{ChetyrkinInt}), for various values of $\alpha$, $\beta$,
${\cal P}_n(p)$ (and also Eq. (3.4) of \cite{Chetyrkin} for diamond
diagrams). Integral $I_c$ of Eq. (\ref{Ic}) can be treated with one
further application of Eq. (\ref{dotprod}), leading to either IR
convergent contributions or tabulated integrals (such as the scalar
``eye'' integral, Eq. (C.5) of \cite{Luscher}). Regarding the integral
$I_d$, employing integration by parts, we find: 
\begin{equation}
I_d = \frac{1}{2}\int \frac{d^4 p\,d^4 k}{(2\pi)^8} 
\frac{\sum_\rho\,\cos (k-q)^\rho}{\hat{p}^2\
  \widehat{p-q}^2\ \hat{k}^2\ \widehat{k-p}^2}
-\frac{1}{2}\sum_\rho\,\partial_{q^\rho}
\int \frac{d^4 p\,d^4 k}{(2
  \pi)^8}\,\frac{(k\, \minuscirc\, q)^\rho}{\hat{p}^2\
  \widehat{p-q}^2\ \hat{k}^2\ \widehat{k-p}^2} 
\label{partialintegration}
\end{equation}
Simple trigonometry on the first term in
Eq. (\ref{partialintegration}): $\sum_\rho\,\cos
(k-q)^\rho=4-\hat{k}^2/2+{\cal O}(q)$ leads to the scalar eye diagram
plus a simple, separable integral. The second term in
Eq. (\ref{partialintegration}) is the derivative of a vector eye
diagram, and can be resolved in two steps: (i) The integrand, $(k\,
\minuscirc\, q)^\rho/(\hat{p}^2\ \widehat{p-q}^2\ \hat{k}^2\
\widehat{k-p}^2)$, after simple trigonometry and use of the symmetry
$k\to p-k$, may be expressed in terms of superficially convergent
and/or known divergent integrals, plus an integrand of the form
$\pcirc^\rho/(\hat{p}^2\ \widehat{p-q}^2\ \hat{k}^2\
\widehat{k-p}^2)$; (ii) the latter, upon contraction with $\qcirc^\rho$
and use of Eq. (\ref{dotprod}), is expressed completely in terms of
known integrals.  

Some superficially convergent integrals, which contain subdivergences,
often appear in various stages of our calculation. A simple prototype
example is: 
$$
\frac{\displaystyle{\sum_\rho}\,(\hat{k}^\rho)^4}{\hat{p}^2\ \widehat{p-q}^2\
  \hat{k}^2\ \widehat{k-p}^2}
$$ 
which is logarithmically divergent for $q\to 0 $. In such cases, a
subtraction of the form: 
$$
\frac{1}{\widehat{k-p}^2}=\frac{1}{\hat{k}^2} +
  \left(\frac{1}{\widehat{k-p}^2} - \frac{1}{\hat{k}^2}\right)
$$
leads to known separable integrals, plus terms in which one can set
$q=0$ without appearance of divergences. 

In conclusion, using the steps which we outlined above, we have
managed to evaluate the integral $I_1$ of 
Eq. (\ref{contract1}), by reducing diamond-like structures into
simpler ones, leading to expressions containing UV-finite integrals
and standard primitively divergent integrals whose values are
known. Using similar considerations, one can also evaluate the
integral $I_2$ of Eq. (\ref{contract2}), which is needed up to order
${\cal   O}(q^2)$. With the evaluation of these two integrals, we can
fully determine the coefficients $A$ and $B$ of Eq. (\ref{ABvalues}),
leading to the calculation of the original two-index integral. Let us
point out that, throughout the whole procedure, the necessity to work
in $D\neq 4$ dimensions does not emerge (It was only necessary in
order to carry out demonstrations, such as Eqs.
(\ref{FirstSubA}-\ref{lastpart2}), leading to the conclusion:
Eq. (\ref{structform4indx})).   

\newpage

\section{Per Diagram Results}

In this appendix we present our perturbative results for the bare
Green's functions, $\Sigma_\Gamma(g_\circ,\,a_{_{\rm L}}\,q)$ (where
$\Gamma=V,\,AV,\,T$), on a per diagram basis. Our results are
expressed in terms of the bare coupling constant, $g_\circ$, the
lattice spacing $a_{_{\rm L}}$, the external momentum, $q$ and the
clover parameter $c_{\rm SW}$. For the sake of 
simplicity we have set $N_c = 3$ in two-loop expressions; at
one-loop level the number of colors is left unspecified and the bare
gauge parameter, $\alpha_\circ$, may take arbitrary values. In all
cases, the number of flavors, $N_f$, can take arbitrary values. 

Only one Feynman diagram, shown in Fig. \ref{ZVAT1loopDiagrams},
contributes to one-loop expressions. Our corresponding results for the
three operators read:
\begin{eqnarray}
\Sigma_{V,\,1-loop}(g_\circ,a_{_{\rm L}}\,q)&=&g_\circ^2\,\frac{\left(N_c^2 -1
  \right)}{N_c}\,\Bigg[\gamma_\mu\,\bigg(-\frac{\alpha_\circ}{32\pi^2}\,\log\left(a^2_{_{\rm
      L}}\,q^2\right)+ 0.0151728775487(3)\,\alpha_\circ \nonumber \\
&&\hspace{3.25cm} + 0.01258087658(1) - 0.007905256548(1)\,c_{\rm SW}
\nonumber \\
&&\hspace{3.25cm} + 0.0027043227859(1)\,c_{\rm SW}^2\bigg) \nonumber \\
&&\hspace{2.25cm} + \frac{q_\mu\,\slashed
    q}{q^2}\left(-\frac{1}{16\pi^2}\,\alpha_\circ\right)\Bigg] 
\label{perdiagramV1loop}  
\end{eqnarray}
\begin{eqnarray}
\Sigma_{AV,\,1-loop}(g_\circ,a_{_{\rm L}}\,q)&=&g_\circ^2\,\frac{\left(N_c^2 -1
  \right)}{N_c}\,\gamma_5\,\Bigg[\gamma_\mu\,\bigg(-\frac{\alpha_\circ}{32\pi^2}\,\log\left(a^2_{_{\rm
      L}}\,q^2\right)+ 0.0151728775487(3)\,\alpha_\circ \nonumber \\
&&\hspace{3.7cm} - 0.002685425493(8) + 0.007905256548(1)\,c_{\rm SW}
\nonumber \\
&&\hspace{3.7cm} - 0.0027043227859(1)\,c_{\rm SW}^2\bigg) \nonumber \\
&&\hspace{2.75cm} + \frac{q_\mu\,\slashed
    q}{q^2}\left(-\frac{1}{16\pi^2}\,\alpha_\circ\right)\Bigg] 
\label{perdiagramAV1loop}
\end{eqnarray}   
\begin{eqnarray}
\Sigma_{T,\,1-loop}(g_\circ,a_{_{\rm L}}\,q)&=&g_\circ^2\,\frac{\left(N_c^2 -1
  \right)}{N_c}\,\gamma_5\,\sigma_{\mu\,\nu}\,\Bigg[\frac{1-\alpha_\circ}{32\pi^2}\,\log\left(a^2_{_{\rm
      L}}\,q^2\right)+ 0.01200659055973(9)\,\alpha_\circ \nonumber \\
&&\hspace{3.65cm} + 0.00118313174(2) -0.0052701710(1)\,c_{\rm SW}
\nonumber \\
&&\hspace{3.65cm} -0.001820704300(1)\,c_{\rm SW}^2\Bigg]
\label{perdiagramT1loop}
\end{eqnarray}   

The contribution to the bare Green's functions from the $\ell$-th
two-loop diagram, can be written in the folowing form:
\begin{eqnarray}
\Sigma^{(\ell)}_{V,\,2-loop}(g_\circ,a_{_{\rm
    L}}\,q)&=&g_\circ^4\,N_f^k\,\Bigg[\gamma_\mu\,
\sum_i c_{\rm SW}^i\Bigg(\frac{{\rm v}_{1,i}^{(\ell)}}{1152\pi^4}\,\log^2\left(a_{_{\rm L}}^2q^2\right) + {\rm v}_{2,i}^{(\ell)}\,\log\left(a_{_{\rm L}}^2q^2\right) +
{\rm v}_{3,i}^{(\ell)}\Bigg) \nonumber \\
&& \hspace{1.1cm} + \frac{q_\mu\,\slashed q}{q^2}\,
\sum_ic_{\rm SW}^i\Bigg(\frac{{\rm v}_{1,i}^{(\ell)}}{288\pi^4}\,\log\left(a_{_{\rm L}}^2q^2\right) + {\rm v}_{4,i}^{(\ell)}\Bigg)\Bigg]
\label{perdiagramV2loop}
\end{eqnarray}

\begin{eqnarray}
\Sigma^{(\ell)}_{AV,\,2-loop}(g_\circ,a_{_{\rm
    L}}\,q)&=&g_\circ^4\,N_f^k\,\gamma_5\,\Bigg[\gamma_\mu
\sum_i c_{\rm SW}^i\Bigg(\frac{w_{1,i}^{(\ell)}}{1152\pi^4}\,\log^2\left(a_{_{\rm L}}^2q^2\right) +
  w_{2,i}^{(\ell)}\,\log\left(a_{_{\rm L}}^2q^2\right) +
  w_{3,i}^{(\ell)}\Bigg) \nonumber \\
&& \hspace{1.6cm} + \frac{q_\mu\,\slashed q}{q^2}\,
\sum_i c_{\rm SW}^i\Bigg(\frac{w_{1,i}^{(\ell)}}{288\pi^4}\,\log\left(a_{_{\rm L}}^2q^2\right) +
w_{4,i}^{(\ell)}\Bigg)\Bigg]
\label{perdiagramAV2loop}
\end{eqnarray}

\begin{equation}
\Sigma^{(\ell)}_{T,\,2-loop}(g_\circ,a_{_{\rm L}}\,q) = 
g_\circ^4\,N_f^k\,\gamma_5\,\sigma_{\mu\,\nu}\,
\Bigg[\sum_i c_{\rm SW}^i\left(\frac{x_{1,i}^{(\ell)}}{1152\pi^4}\,\log^2\left(a_{_{\rm L}}^2q^2\right) +
  x_{2,i}^{(\ell)}\,\log\left(a_{_{\rm L}}^2q^2\right) +
  x_{3,i}^{(\ell)}\right)\Bigg]
\label{perdiagramT2loop}
\end{equation}
where the index $\ell$ runs over all contributing two-loop
diagrams. The dependence on $c_{\rm SW}$ is polynomial of degree up to
4 ($i=0,\cdots,\,4$). The number of flavors, $N_f$\,, is raised to the
power $k$ where, of course, $k=1$ only for diagrams 8 and 9
of Fig. \ref{ZVAT2loopDiagrams}, since they are the only diagrams
containing a closed fermion loop; the remaining diagrams have
$k=0$. The coefficients ${\rm v}^{(\ell)}$, 
$w^{(\ell)}$ and $x^{(\ell)}$ are numerical constants obtained upon
evaluating each two-loop Feynman diagram. In Tables
\ref{tabVa}-\ref{tabVb}, \ref{tabAVa}-\ref{tabAVb} and
\ref{tabTa}-\ref{tabTb}, we present our results for ${\rm
  v}^{(\ell)}$, $w^{(\ell)}$ and $x^{(\ell)}$, respectively, with
accuracy up to 10 decimal places. 

In the case of the singlet Axial-Vector operator, it turns out that
the only additional diagram contributing to
$\Sigma^{(\ell)}_{AV,singlet}(a_{_{\rm L}}q)$ is diagram 4 of
Fig. \ref{singletVAT}. It is a straight forward exercise to recover
the bare matrix element; starting from
$Z_{AV,\,singlet}^{L,RI^{\prime}}$ (Eq. (\ref{ScalarSingletRI}) and
$Z_{\psi}^{L,RI^{\prime}}$ (Eq. (\ref{Zpsi2loopRI}), one can employ the $RI^\prime$
renormalization condition, Eq. \ref{ZXrulesb}, to extract the corresponding matrix
element.  

\begin{table}
\begin{center}
\caption{Contribution of two-loop diagrams to
  $\Sigma^{(\ell)}_{V,2-loop}$\ \  ($\ell=1{-}14$)
\label{tabVa}}
\begin{tabular}{ccc@{}cr@{}lr@{}lr@{}l}
\multicolumn{1}{c}{$\ell$}&
\multicolumn{1}{c}{$i$}&
\multicolumn{2}{c}{${\rm v}^{(\ell)}_{1,i}$} &
\multicolumn{2}{c}{${\rm v}^{(\ell)}_{2,i}$} &
\multicolumn{2}{c}{${\rm v}^{(\ell)}_{3,i}$} &
\multicolumn{2}{c}{${\rm v}^{(\ell)}_{4,i}$} \\
\hline \hline
      & 0 & &0& 0&.0010901413    & -0&.00955549(5) & 0&.0021802826   \\
1     & 1 & &0& 0&               &  0&.0089908(2)  & 0& \\
      & 2 & &0& 0&               & -0&.00522028(9) & 0& \\
\hline
      & 0 & &0& 0&&  0&.00230566(3)  & 0& \\   
2     & 1 & &0& 0&&  0&              & 0& \\
      & 2 & &0& 0&& -0&.000041360(2) & 0& \\
\hline                                           
       & 0 & &0& -0&.0004010149       &  0&.01183022(2) & -0&.0008020299 \\
3-7   & 1 & &0&  0&                  & -0&.00738844(3) &  0& \\
      & 2 & &0&  0&                  &  0&.00260794(1) &  0& \\
\hline
      & 0 & &0& 0&.0000534687       & -0&.0003947(1)   & 0&.0001069374 \\
      & 1 & &0& 0&                  &  0&.00032545(2)  & 0& \\                
8-9   & 2 & &0& 0&                  & -0&.00077830(1)  & 0& \\ 
      & 3 & &0& 0&                  &  0&.000389303(9) & 0& \\
      & 4 & &0& 0&                  & -0&.000146063(2) & 0& \\
\hline
       & 0 & &0& -0&.0017442261     &  0&.0152888292        & -0&.0034884521 \\ 
 $\ $10-11$\ $ & 1 & &0&  0&                & -0&.0043548025        &  0& \\
       & 2 & &0&  0&                &  0&.0014897419        &  0& \\
\hline    
      & 0 & &0& 0&&  0&              & 0& \\
12    & 1 & &0& 0&&  0&.000063005(3) & 0& \\
      & 2 & &0& 0&& -0&.000061816(2) & 0& \\
\hline
       & 0 & &0&  0&.0000390267        & -0&.000161967(5) &  0&.0000780534  \\ 
 13    & 1 & &0&  0&.0001255891        & -0&.00099763(1)  &  0&.0002511781  \\
       & 2 & &0& -0&.0002000264        &  0&.00009461(1)  & -0&.0004000527  \\
       & 3 & &0&  0&                   & -0&.000046762(1) &  0&                   \\
\hline
      & 0 & &0& 0&&  0&.00045044(3)  &  0& \\
14    & 1 & &0& 0&& -0&.00147021(3)  &  0& \\
      & 2 & &0& 0&& -0&.000043209(1) &  0& \\
      & 3 & &0& 0&&  0&.000057664(2) &  0& \\
\hline\hline
\end{tabular}
\end{center}
\end{table}

\begin{table}
\begin{center}
\caption{Contribution of two-loop diagrams to
  $\Sigma^{(\ell)}_{V,2-loop}$\ \  ($\ell=15{-}20$)
\label{tabVb}}
\begin{tabular}{ccr@{}lr@{}lr@{}lr@{}l}
\multicolumn{1}{c}{$\ell$}&
\multicolumn{1}{c}{$i$}&
\multicolumn{2}{c}{${\rm v}^{(\ell)}_{1,i}$} &
\multicolumn{2}{c}{${\rm v}^{(\ell)}_{2,i}$} &
\multicolumn{2}{c}{${\rm v}^{(\ell)}_{3,i}$} &
\multicolumn{2}{c}{${\rm v}^{(\ell)}_{4,i}$} \\
\hline \hline
       & 0 & 27&& -0&.0020904154        &  0&.0061629(1)   & -0&.00418077(4)       \\
 $\ $15$\ $    & 1 &  0&& -0&.0005507191        & -0&.0010101(1)   & -0&.0011014381  \\
       & 2 &  0&&  0&.0001759227        &  0&.00026474(6)  &  0&.0003518455 \\
       & 3 &  0&&  0&                   &  0&.000424491(6) &  0&                   \\
\hline
       & 0 & 0&&  0&.0000390267        & -0&.000342083(4) &  0&.0000780534  \\
 16    & 1 & 0&&  0&.0001255891        &  0&.001157216(6) &  0&.0002511781 \\
       & 2 & 0&& -0&.0002000264        &  0&.00178325(2)  & -0&.0004000527 \\
       & 3 & 0&&  0&                   & -0&.000076152(1) &  0&                   \\
\hline
       & 0 & 0&& -0&.0000356458  & -0&.000049925(1)   & 0&.00001874(2) \\
       & 1 & 0&&  0&             &  0&.000198131(2)   & 0& \\
17     & 2 & 0&&  0&             & -0&.000064731(4)   & 0& \\
       & 3 & 0&&  0&             & -0&.0000161812(2)  & 0& \\
       & 4 & 0&&  0&             &  0&.0000032674     & 0& \\
\hline
       & 0 & -1&&  0&.0000558195         &  0&.000098634(7)  &  0&.00002159(5)        \\
       & 1 &  0&& -0&.0000211097         & -0&.000057416(3)  & -0&.0000422194 \\
 18    & 2 &  0&&  0&.0000249033         & -0&.000179175(1)  &  0&.0000498067 \\
       & 3 &  0&&  0&                    &  0&.0000335005(3) &  0&                    \\
       & 4 &  0&&  0&                    &  0&.0000077358(2) &  0&                    \\
\hline
       & 0 & -8&& 0&.0002681483 & -0&.0086175(2)  & -0&.0000340358 \\
       & 1 &  0&& 0&.0003206506 &  0&.0144022(6)  &  0&.0006413012 \\
 19    & 2 &  0&& 0&.0001992266 &  0&.00454038(9) &  0&.0003984533 \\
       & 3 &  0&& 0&          &  0&.00128588(2) &  0&          \\
       & 4 &  0&& 0&          & -0&.00047870(1) &  0&          \\
\hline
       & 0 & 4&& -0&.0008031274        &  0&.00453973(4)   & -0&.00103589(2)       \\
       & 1 & 0&&  0&.0001779933        & -0&.00234305(6)   &  0&.0003559866 \\
 20    & 2 & 0&& -0&.0000608900        &  0&.00090217(4)   & -0&.0001217801 \\
       & 3 & 0&&  0&                   &  0&.000065108(4)  &  0&                  \\
       & 4 & 0&&  0&                   &  0&.0000428543(8) &  0&                  \\
\hline\hline
\end{tabular}
\end{center}
\end{table}

\begin{table}
\begin{center}
\caption{Contribution of two-loop diagrams to
  $\Sigma^{(\ell)}_{AV,2-loop}$\ \  ($\ell=1{-}14$)
\label{tabAVa}}
\begin{tabular}{ccc@{}cr@{}lr@{}lr@{}l}
\multicolumn{1}{c}{$\ell$}&
\multicolumn{1}{c}{$i$}&
\multicolumn{2}{c}{$w^{(\ell)}_{1,i}$} &
\multicolumn{2}{c}{$w^{(\ell)}_{2,i}$} &
\multicolumn{2}{c}{$w^{(\ell)}_{3,i}$} &
\multicolumn{2}{c}{$w^{(\ell)}_{4,i}$} \\
\hline \hline
       & 0 & &0& 0&.0010901413    & -0&.004299380(2) & 0&.0021802826 \\
1      & 1 & &0& 0&               & -0&.0089908(2)   & 0& \\
       & 2 & &0& 0&               &  0&.00522028(9)  & 0& \\
\hline
      & 0 & &0& 0&& -0&.00010270(1)  & 0& \\   
2     & 1 & &0& 0&&  0&              & 0& \\
      & 2 & &0& 0&&  0&.000041360(2) & 0& \\
\hline                                           
      & 0 & &0& -0&.0004010149       & -0&.00177094(2) & -0&.0008020299 \\
3-7   & 1 & &0&  0&                  &  0&.00738844(3) &  0& \\
      & 2 & &0&  0&                  & -0&.00260794(1) &  0& \\
\hline
      & 0 & &0& 0&.0000534687       & -0&.0000143(1)   & 0&.0001069374 \\
      & 1 & &0& 0&                  & -0&.00019712(2)  & 0& \\                
8-9   & 2 & &0& 0&                  &  0&.00001658(1)  & 0& \\ 
      & 3 & &0& 0&                  & -0&.000389303(9) & 0& \\
      & 4 & &0& 0&                  &  0&.000146063(2) & 0& \\
\hline
       & 0 & &0& -0&.0017442261     &  0&.0068790161        & -0&.0034884521 \\ 
 $\ $10-11$\ $ & 1 & &0&  0&                &  0&.0043548025        &  0& \\
       & 2 & &0&  0&                & -0&.0014897419        &  0& \\
\hline    
      & 0 & &0& 0&&  0&              & 0& \\
12    & 1 & &0& 0&& -0&.000063005(3) & 0& \\
      & 2 & &0& 0&&  0&.000061816(2) & 0& \\
\hline
       & 0 & &0&  0&.0000390267        &  0&.00017234(2)  &  0&.0000780534  \\ 
 13    & 1 & &0&  0&.0001255891        &  0&.00172937(5)  &  0&.0002511781 \\
       & 2 & &0& -0&.0002000264        & -0&.00147678(5)  & -0&.0004000527 \\
       & 3 & &0&  0&                   &  0&.000046762(1) &  0&                   \\
\hline
      & 0 & &0& 0&&  0&.000203052(3) &  0& \\
14    & 1 & &0& 0&&  0&.00057021(3)  &  0& \\
      & 2 & &0& 0&&  0&.00010067(1)  &  0& \\
      & 3 & &0& 0&& -0&.000057664(2) &  0& \\
\hline\hline
\end{tabular}
\end{center}
\end{table}

\begin{table}
\begin{center}
\caption{Contribution of two-loop diagrams to
  $\Sigma^{(\ell)}_{AV,2-loop}$\ \  ($\ell=15{-}20$)
\label{tabAVb}}
\begin{tabular}{ccr@{}lr@{}lr@{}lr@{}l}
\multicolumn{1}{c}{$\ell$}&
\multicolumn{1}{c}{$i$}&
\multicolumn{2}{c}{$w^{(\ell)}_{1,i}$} &
\multicolumn{2}{c}{$w^{(\ell)}_{2,i}$} &
\multicolumn{2}{c}{$w^{(\ell)}_{3,i}$} &
\multicolumn{2}{c}{$w^{(\ell)}_{4,i}$} \\
\hline \hline
       & 0 & 27&& -0&.0020904154        &  0&.0019129(1)   & -0&.00418077(4)       \\
 $\ $15$\ $    & 1 &  0&& -0&.0005507191        &  0&.0048214(1)   & -0&.0011014381  \\
       & 2 &  0&&  0&.0001759227        & -0&.00100389(6)  &  0&.0003518455 \\
       & 3 &  0&&  0&                   & -0&.000424491(6) &  0&                   \\
\hline
       & 0 & 0&&  0&.0000390267        & -0&.0001539166(3) &  0&.0000780534  \\
 16    & 1 & 0&&  0&.0001255891        &  0&.00125957(2)   &  0&.0002511781 \\
       & 2 & 0&& -0&.0002000264        &  0&.001151401(6)  & -0&.0004000527 \\
       & 3 & 0&&  0&                   &  0&.000076152(1)  &  0&                   \\
\hline
       & 0 & 0&& -0&.0000356458  & -0&.000281627(7)   & 0&.0005068330 \\
       & 1 & 0&&  0&             &  0&.000060887(2)   & 0& \\
 17    & 2 & 0&&  0&             &  0&.000050595(4)   & 0& \\
       & 3 & 0&&  0&             & -0&.000006587(2)   & 0& \\
       & 4 & 0&&  0&             & -0&.0000046598(4)  & 0& \\
\hline
       & 0 & -1&&  0&.0000558195         &  0&.000050308(7)  &  0&.00002159(5)        \\
       & 1 &  0&& -0&.0000211097         & -0&.000226495(3)  & -0&.0000422194 \\
 18    & 2 &  0&&  0&.0000249033         &  0&.000086519(1)  &  0&.0000498067 \\
       & 3 &  0&&  0&                    & -0&.0000336128(3) &  0&                    \\
       & 4 &  0&&  0&                    & -0&.0000077358(2) &  0&                    \\
\hline
       & 0 & -8&& 0&.0002681483 & -0&.0206450(5)   & -0&.0000340358 \\
       & 1 &  0&& 0&.0003206506 &  0&.0253960(7)   &  0&.0006413012 \\
 19    & 2 &  0&& 0&.0001992266 &  0&.0064762(2)   &  0&.0003984533 \\
       & 3 &  0&& 0&          & -0&.00199356(7)  &  0&          \\
       & 4 &  0&& 0&          & -0&.000442050(6) &  0&          \\
\hline
       & 0 & 4&& -0&.0004593941        &  0&.00110429(4)   & -0&.00034842(2)       \\
       & 1 & 0&& -0&.0001779933        &  0&.00063376(6)   & -0&.0003559866 \\
 20    & 2 & 0&&  0&.0000608900        & -0&.00029572(4)   &  0&.0001217801 \\
       & 3 & 0&&  0&                   & -0&.000201975(4)  &  0&                  \\
       & 4 & 0&&  0&                   &  0&.0000060939(8) &  0&                  \\
\hline\hline
\end{tabular}
\end{center}
\end{table}

\begin{table}
\begin{flushleft}
\phantom{a}\vspace{-0.5cm}
\begin{minipage}[h]{0.45\linewidth}
\caption{Contribution of two-loop\\
diagrams to $\Sigma^{(\ell)}_{T,2-loop}$\ \  ($\ell=1{-}14$)
\label{tabTa}}
\begin{tabular}{ccr@{}lr@{}lr@{}l}
\multicolumn{1}{c}{$\ell$}&
\multicolumn{1}{c}{$i$}&
\multicolumn{2}{c}{$x^{(\ell)}_{1,i}$} &
\multicolumn{2}{c}{$x^{(\ell)}_{2,i}$} &
\multicolumn{2}{c}{$x^{(\ell)}_{3,i}$} \\
\hline \hline
      & 0 & 0&& 0&& -0&.0045414(8) \\
1     & 1 & 0&& 0&&  0&.0059942(3) \\
      & 2 & 0&& 0&&  0&.0035149(4) \\
\hline
      & 0 & 0&& 0&&  0&.0010537(2) \\
2     & 1 & 0&& 0&&  0&            \\
      & 2 & 0&& 0&& -0&.0012514(2) \\
\hline                                           
       & 0 & -15&& 0&.0028583491      &  0&.001177(1)    \\
3-7    & 1 &   0&& 0&                 & -0&.00492566(5)  \\
       & 2 &   0&& 0&                 & -0&.001583088(1) \\
\hline
       & 0 & 2&& -0&.0000803010      & -0&.0000461(3)   \\
       & 1 & 0&&  0&.0000426116      &  0&.00022585(9)  \\
 8-9   & 2 & 0&& -0&.0002519813      & -0&.00021615(4)  \\
       & 3 & 0&&  0&                 &  0&.00024528(3)  \\
       & 4 & 0&&  0&                 &  0&.000058928(5) \\
\hline
       & 0 & 0&&  0&&  0&.0072658787  \\
 $\ $10-11$\ $ & 1 & 0&&  0&& -0&.0029032016  \\
       & 2 & 0&&  0&& -0&.0010029792 \\
\hline    
      & 0 & 0&& 0&&  0&               \\
12    & 1 & 0&& 0&&  0&.0003050(1)    \\
      & 2 & 0&& 0&&  0&.0000039204(9) \\
\hline
      & 0 & 0&&  0&& -0&.00002584(1)  \\
13    & 1 & 0&&  0&&  0&.00029039(5)  \\
      & 2 & 0&&  0&& -0&.0008905(1)   \\
      & 3 & 0&&  0&& -0&.000030105(2) \\
\hline
       & 0 & 0&& 0&&  0&.0002775(1)   \\
 14    & 1 & 0&& 0&& -0&.0015909(4)   \\
       & 2 & 0&& 0&&  0&.00030041(3)  \\
       & 3 & 0&& 0&& -0&.000030400(2) \\
\hline\hline
\end{tabular}
\end{minipage}
\end{flushleft}
\phantom{a}\vspace{-25.25cm}
\begin{flushright}
\begin{minipage}[h]{0.45\linewidth}
\caption{Contribution of two-loop\\
 diagrams to $\Sigma^{(\ell)}_{T,2-loop}$\ \  ($\ell=15{-}20$)
\label{tabTb}}
\begin{tabular}{ccc@{}cr@{}lr@{}l}
\multicolumn{1}{c}{$\ell$}&
\multicolumn{1}{c}{$i$}&
\multicolumn{2}{c}{$x^{(\ell)}_{1,i}$} &
\multicolumn{2}{c}{$x^{(\ell)}_{2,i}$} &
\multicolumn{2}{c}{$x^{(\ell)}_{3,i}$} \\
\hline \hline
      & 0 & &0& 0&&  0&.0017380(8)   \\
 $\ $15$\ $    & 1 & &0& 0&& -0&.0011468(3)   \\
      & 2 & &0& 0&& -0&.00090530(9)  \\
      & 3 & &0& 0&&  0&.000054081(9) \\
\hline
      & 0 & &0&  0&& -0&.00016259(4)  \\
16    & 1 & &0&  0&&  0&.0007288(1)   \\
      & 2 & &0&  0&&  0&.0007569(1)   \\
      & 3 & &0&  0&& -0&.000050771(5) \\
\hline
      & 0 & &0&  0&.0000356458       &  0&.00003599(7)   \\
      & 1 & &0&  0&                  &  0&.000034379(6)  \\
17    & 2 & &0&  0&                  & -0&.00007614(2)   \\
      & 3 & &0&  0&                  & -0&.000023760(4)  \\
      & 4 & &0&  0&                  & -0&.0000109914(9) \\
\hline
      & 0 & &0&  0&.0000356457 &  0&.0001341(1)    \\
      & 1 & &0&  0&            & -0&.00022585(1)   \\
18    & 2 & &0&  0&            & -0&.00003143(2)   \\
      & 3 & &0&  0&            &  0&.0000069846(4) \\
      & 4 & &0&  0&            & -0&.0000045409(4) \\
\hline
      & 0 & &0& 0&&  0&.026645(7)   \\
      & 1 & &0& 0&& -0&.04180(1)    \\
19    & 2 & &0& 0&& -0&.00994(5)    \\
      & 3 & &0& 0&&  0&.0049633(8)  \\
      & 4 & &0& 0&&  0&.00067162(8) \\
\hline
      & 0 & &0&  0&&  0&.0010004(2)    \\
      & 1 & &0&  0&& -0&.0071228(5)    \\
20    & 2 & &0&  0&& -0&.00010774(3)   \\
      & 3 & &0&  0&&  0&.000068781(9)  \\
      & 4 & &0&  0&&  0&.0000163351(2) \\
\hline\hline
\end{tabular}
\end{minipage}
\end{flushright}
\end{table}

\begin{acknowledgments}
This work is supported in part by the
Research Promotion Foundation of Cyprus 
(Proposal Nr: $\rm ENI\Sigma X$/0506/17). 
\end{acknowledgments}


\newpage
\phantom{a}

\vskip 4.5cm
\begin{figure}[h]
\centerline{\psfig{figure=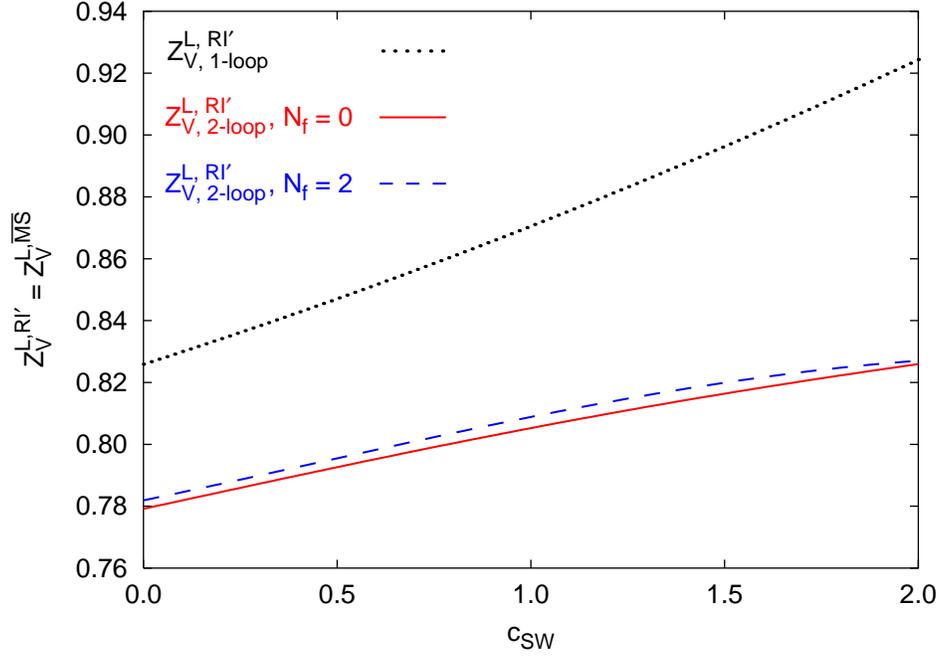,scale=0.50, angle=-90}}
\caption{$Z_V^{L,RI'}(a_{_{\rm L}}\bar{\mu})=Z_V^{L,\overline{MS}}(a_{_{\rm L}}\bar{\mu})$ versus $c_{{\rm SW}}$ 
($N_c=3$, $\bar{\mu}=1/a_{_{\rm L}}$, $\beta_{\rm o}=6.0$). Results up to 2 loops are shown for $N_f=0$ (solid line) and $N_f=2$ (dashed line); one-loop results are plotted with a dotted line.  \label{ZVRIplot}}
\end{figure}
\newpage
\begin{figure}[p]
\centerline{\psfig{figure=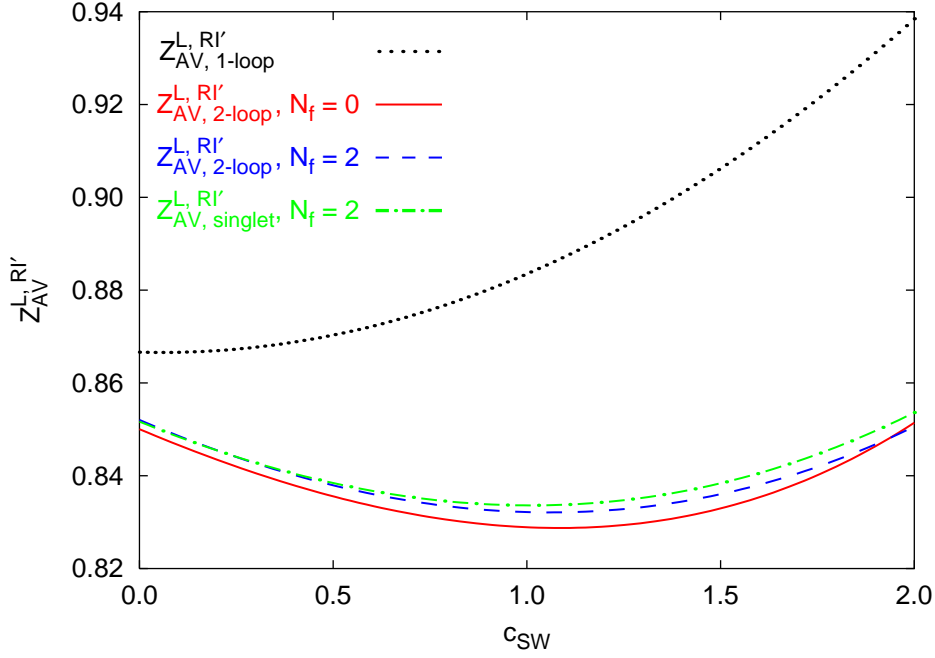,scale=0.50, angle=-90}}
\caption{$Z_{AV}^{L,RI'}(a_{_{\rm L}}\bar{\mu})$ versus $c_{{\rm SW}}$ 
($N_c=3$, $\bar{\mu}=1/a_{_{\rm L}}$, $\beta_{\rm o}=6.0$). 
Results up to 2 loops, for the flavor nonsinglet operator, 
are shown for $N_f=0$ (solid line) and $N_f=2$ (dashed line); 
2-loop results for the flavor singlet operator, for $N_f=2$, 
are plotted with a dash-dotted line; one-loop results are 
plotted with a dotted line. \label{ZARIplot}}
\end{figure}
\begin{figure}[p]
\centerline{\psfig{figure=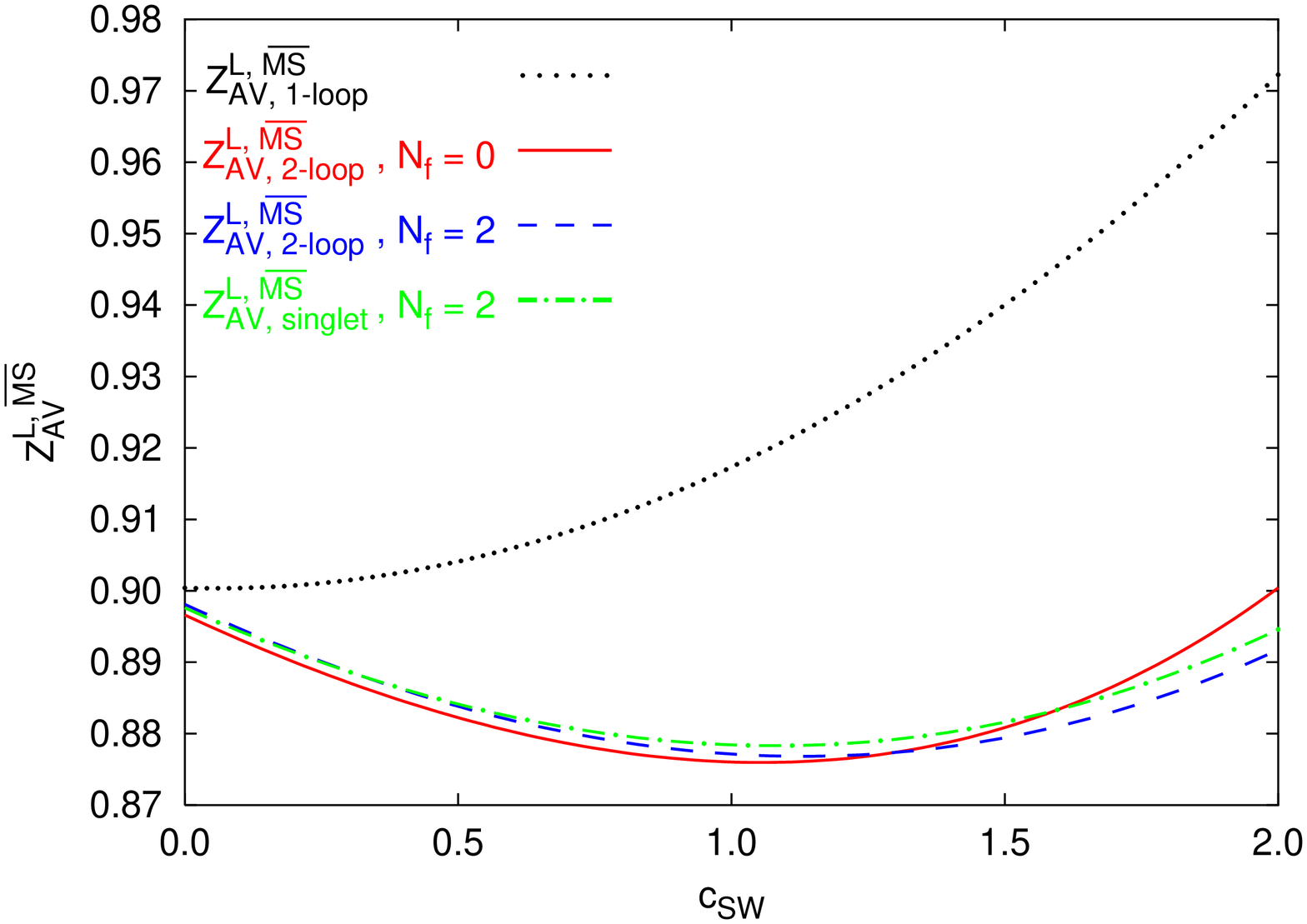,scale=0.50, angle=-90}}
\caption{$Z_{AV}^{L,\overline{MS}}(a_{_{\rm L}}\bar{\mu})$ versus $c_{{\rm SW}}$ 
($N_c=3$, $\bar{\mu}=1/a_{_{\rm L}}$, $\beta_{\rm o}=6.0$). 
Same notation as in Fig. \ref{ZARIplot}. \label{ZAMSplot}}
\end{figure}
\begin{figure}[p]
\centerline{\psfig{figure=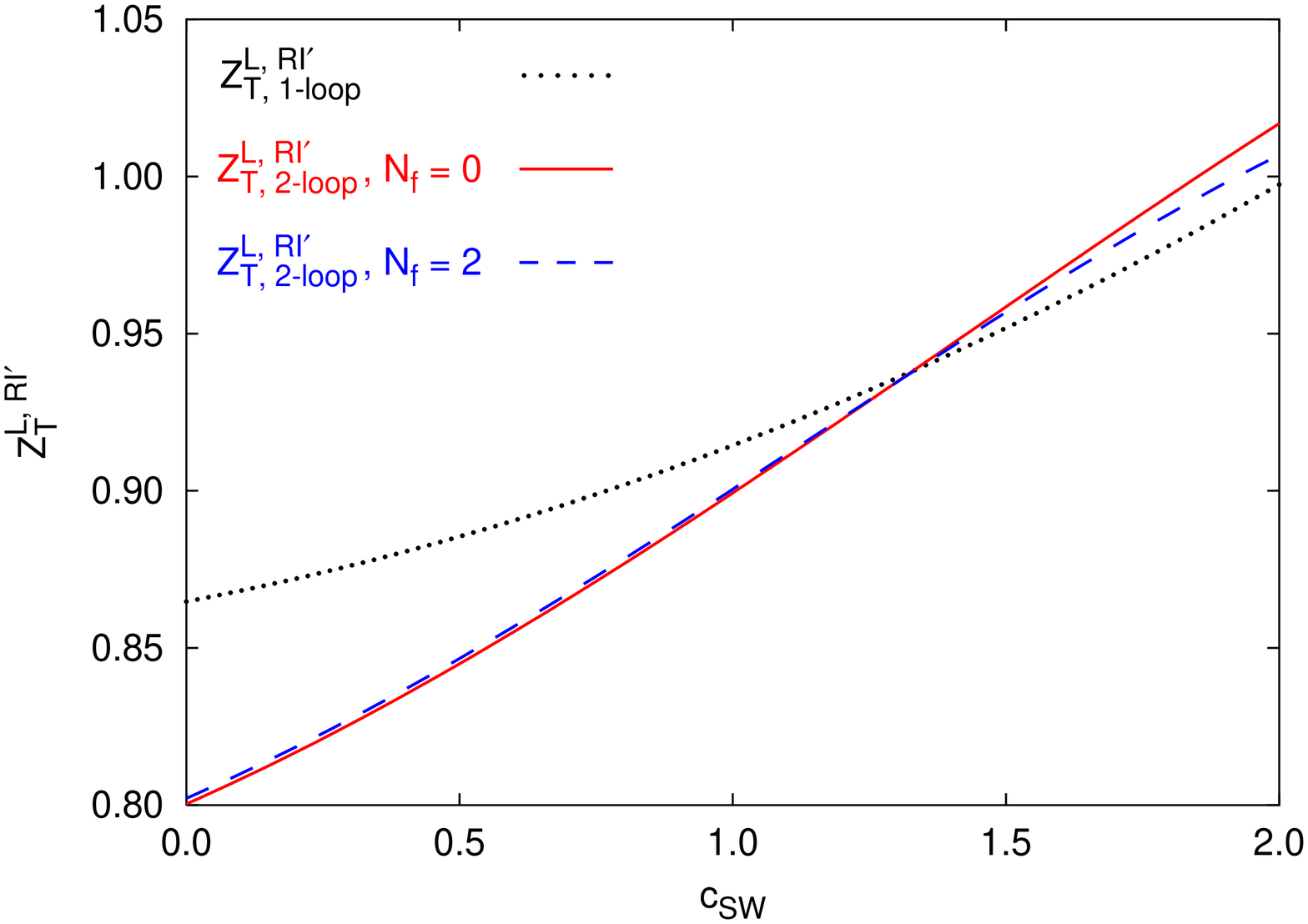,scale=0.50, angle=-90}}
\caption{$Z_T^{L,RI'}(a_{_{\rm L}}\bar{\mu})$ versus $c_{{\rm SW}}$ 
($N_c=3$, $\bar{\mu}=1/a_{_{\rm L}}$, $\beta_{\rm o}=6.0$). Results up to 2 loops are shown for $N_f=0$ (solid line) and $N_f=2$ (dashed line); one-loop results are plotted with a dotted line. \label{ZTRIplot}}
\end{figure}
\begin{figure}[p]
\centerline{\psfig{figure=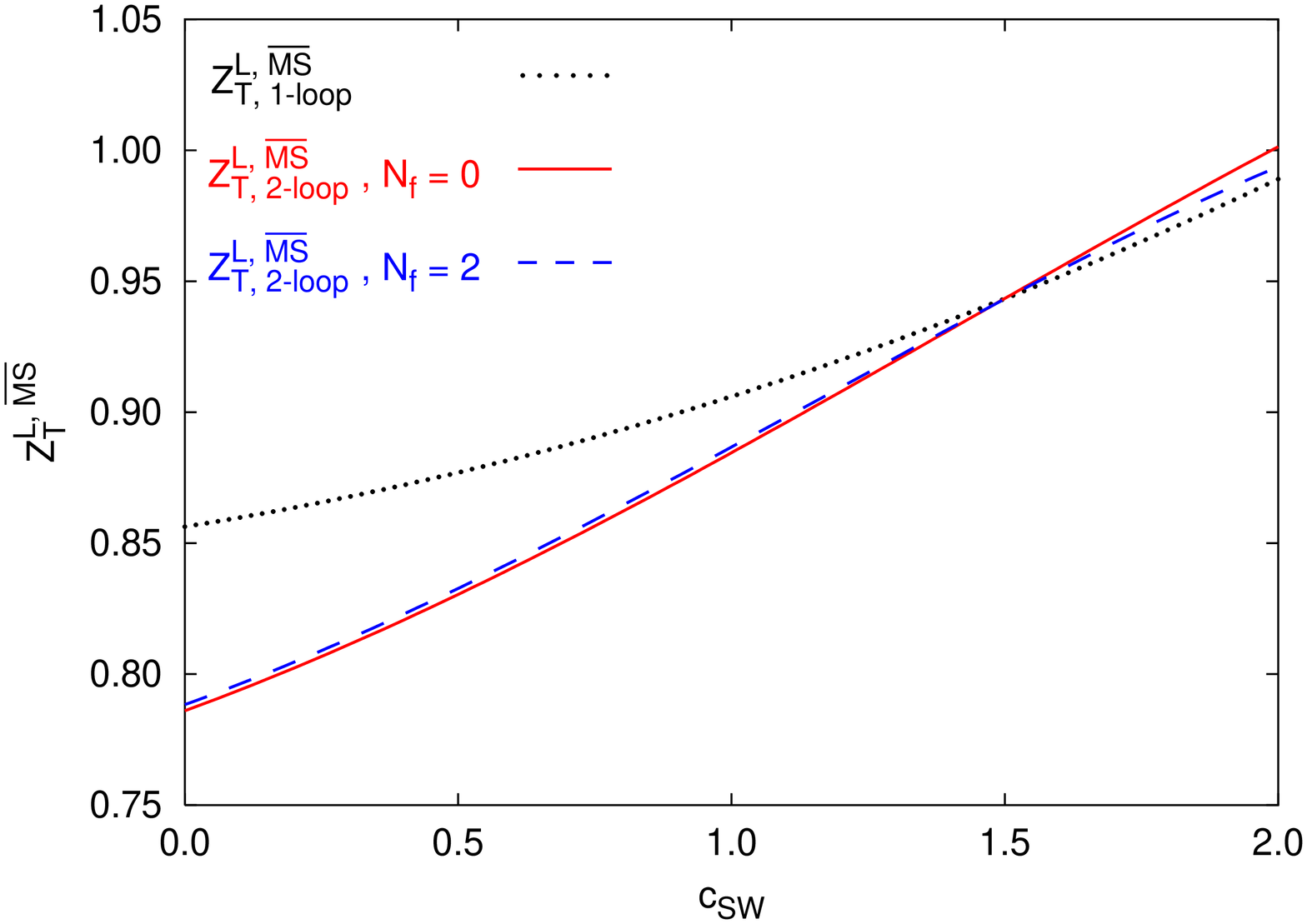,scale=0.50, angle=-90}}
\caption{$Z_T^{L,\overline{MS}}(a_{_{\rm L}}\bar{\mu})$ versus $c_{{\rm SW}}$ 
($N_c=3$, $\bar{\mu}=1/a_{_{\rm L}}$, $\beta_{\rm o}=6.0$). Same
  notation as in Fig. \ref{ZTRIplot}. \label{ZTMSplot}}
\end{figure}
\begin{figure}[p]
\centerline{\psfig{figure=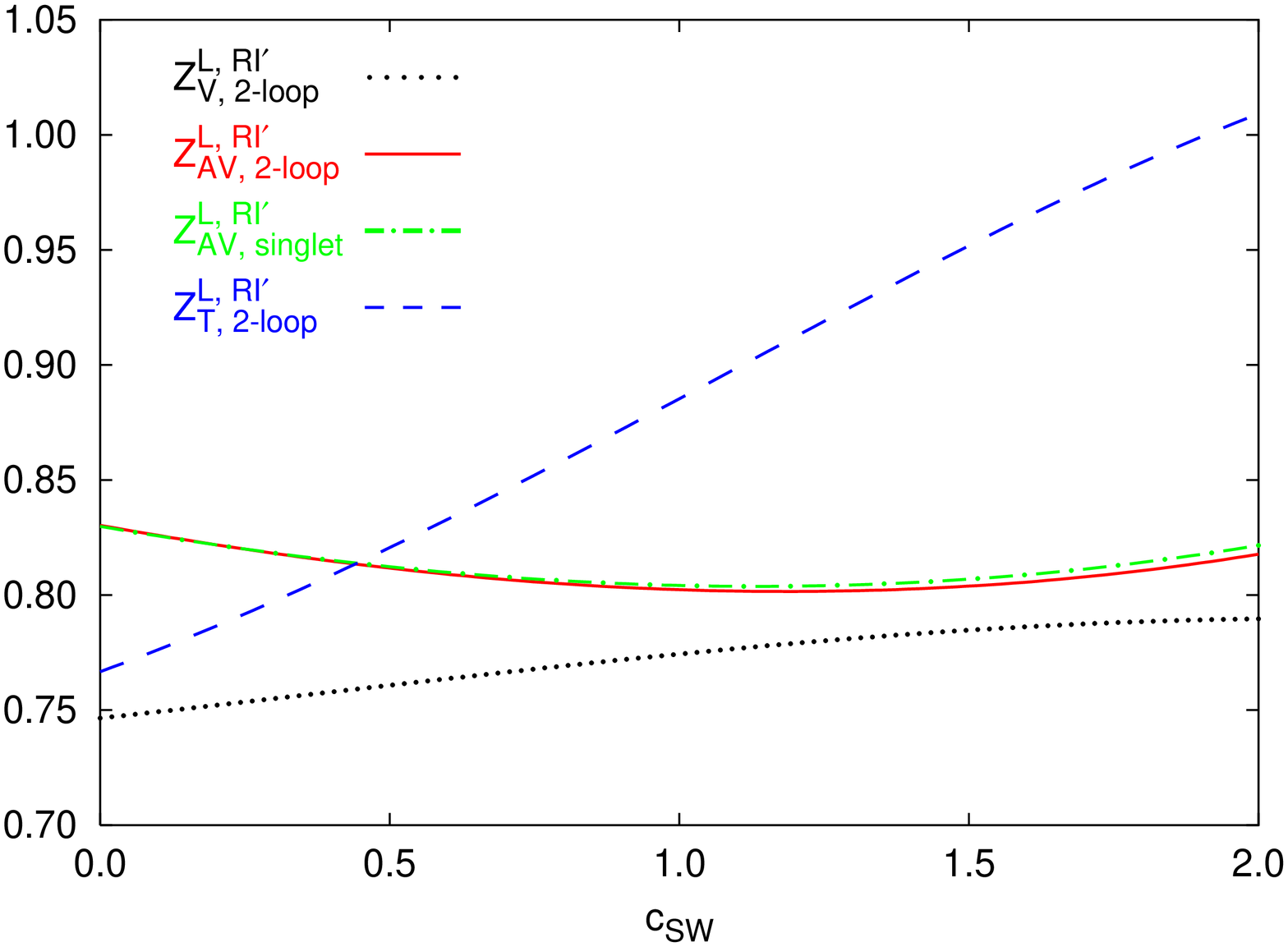,scale=0.50, angle=-90}}
\caption{\Black{$Z_V^{L,RI^{\prime}}$} (dotted line), 
\Red{$Z_{AV}^{L,RI^{\prime}}$} (solid line), 
\Green{$Z_{AV, singlet}^{L,RI^{\prime}}$} (dash-dotted line) and
\Blue{$Z_T^{L,RI^{\prime}}$} (dashed line)
up to 2 loops, versus $c_{\rm SW}$ ($N_c=3$, $\bar{\mu}=1/a_{_{\rm L}}$, 
$N_f=2$, $\beta_{\rm o}=5.3$). \label{ZVATRIplot}}
\end{figure}
\begin{figure}[p]
\centerline{\psfig{figure=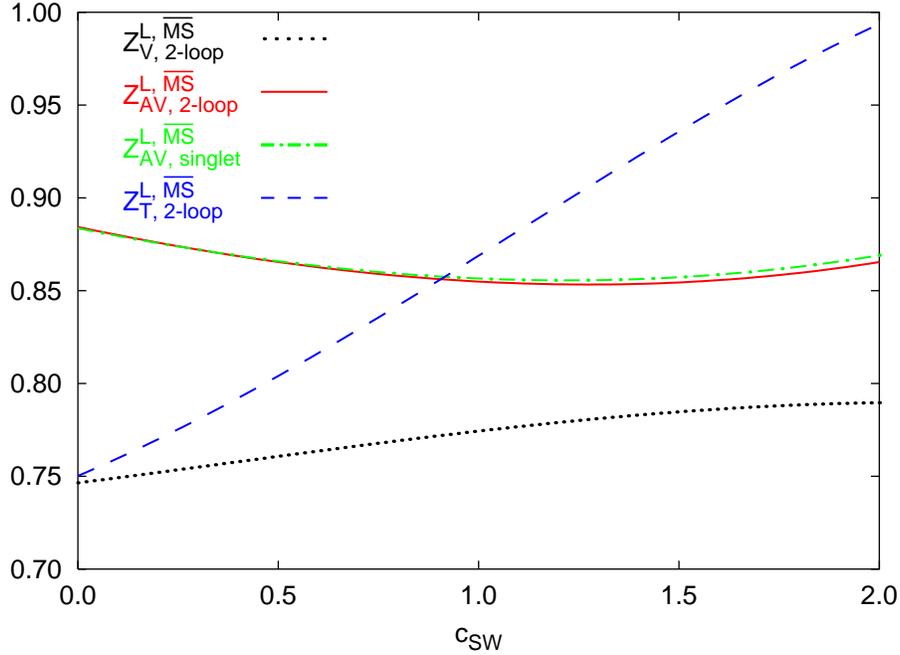,scale=0.50, angle=-90}}
\caption{\Black{$Z_V^{L,\overline{MS}}$} (dotted line), 
\Red{$Z_{AV}^{L,\overline{MS}}$} (solid line), 
\Green{$Z_{AV, singlet}^{L,\overline{MS}}$} (dash-dotted line) and
\Blue{$Z_T^{L,\overline{MS}}$} (dashed line)
up to 2 loops, versus $c_{\rm SW}$ ($N_c=3$, $\bar{\mu}=1/a_{_{\rm L}}$, 
$N_f=2$, $\beta_{\rm o}=5.3$). \label{ZVATMSplot}}
\end{figure}

\end{document}